\documentclass[notitlepage, hidelinks,12pt]{article}
\usepackage[square]{natbib}
\usepackage{graphicx}
\usepackage{appendix}
\usepackage{color}
\usepackage[gen]{eurosym}
\usepackage{fullpage}
\usepackage{hyperref}
\usepackage{setspace}
\usepackage{amsmath}
\usepackage[multiple]{footmisc}
\usepackage{lipsum}
\usepackage{chngcntr}

\begin{document}

\title{Lockdown Strategies, Mobility Patterns and COVID-19\thanks{We thank an anonymous reviewer, Jan van Ours and Adrian Nieto Castro for discussions. $\bullet$ Askitas: askitas@iza.org $\bullet$ Tatsiramos: konstantinos.tatsiramos@uni.lu $\bullet$ Verheyden: bertrand.verheyden@liser.lu}}

\author{\rule[15pt]{0pt}{15pt}Nikos Askitas \\
\small{IZA - Institute of Labor Economics and CESifo} \\
\and
Konstantinos Tatsiramos  \\
\small{University of Luxembourg, LISER, IZA and CESifo} \\
\and 
Bertrand Verheyden  \\
\small{Luxembourg Institute of Socio-Economic Research (LISER)} 
}
\date{\normalsize{\rule[13pt]{0pt}{13pt} May 27, 2020}}


\begin{titlepage}
\maketitle
\thispagestyle{empty}
\setcounter{page}{0}

\begin{abstract} 
\singlespacing

\small{\noindent
We develop a multiple-events model and exploit within and between country variation in the timing, type and level of intensity of various public policies to study their dynamic effects on the daily incidence of COVID-19 and on population mobility patterns across 135 countries. We remove concurrent policy bias by taking into account the contemporaneous presence of multiple interventions. The main result of the paper is that cancelling public events and imposing restrictions on private gatherings followed by school closures have quantitatively the most pronounced effects on reducing the daily incidence of COVID-19. They are followed by workplace as well as stay-at-home requirements, whose statistical significance and levels of effect are not as pronounced. Instead, we find no effects for international travel controls, public transport closures and restrictions on movements across cities and regions. We establish that these findings are mediated by their effect on population mobility patterns in a manner consistent with time-use and epidemiological factors.}
\bigskip

\noindent \normalsize{\textbf{\rule[10pt]{0pt}{10pt}Keywords:} COVID-19, public policies, non-pharmaceutical interventions, multiple events, mobility}

\noindent \normalsize{\textbf{J\rule[10pt]{0pt}{10pt}EL codes:} I12, I18, G14}

\end{abstract}
\end{titlepage}

\newpage
\doublespacing

\section{Introduction}\label{section:intro}
\noindent
In December 2019, the COVID-19 outbreak was registered in Wuhan China. The World Health Organization declared it a ``Public Health Emergency of International Concern" on January 30, 2020 and escalated it to a pandemic on March 11, 2020. The disease has been recorded in over 200 countries and territories with several millions of confirmed cases and a case mortality rate of around seven percent.\footnote{\href{https://coronavirus.jhu.edu/map.html}{COVID-19 Dashboard by the CSSE at Johns Hopkins University (JHU).}} In the early stages of the outbreak, attempts were made to trace every infection back to its origin. Tracing back to the ``index'' case on an international level soon became impossible and most countries responded by imposing restrictions on international travel. In the later stages of the epidemic, a number of non - pharmaceutical interventions (henceforth referred to as NPIs or public policies) were undertaken, which were of a domestic nature revolving around the idea of ``social distancing''. The aim of these interventions was to slow down the pandemic by restricting mobility so that it does not overwhelm health system capacities.

This paper studies how lockdown policies affect the daily incidence of COVID-19 and population mobility patterns across 135 countries based on several data sources.\footnote{The data sources include: i) Coded government response data obtained from \cite{hale2020oxford}, ii) prevalence data from European Centre for Disease Prevention and Control (ECDC) and iii) population mobility data from Google Community Mobility Reports. The analysis includes the latest data up to this writing.}  Understanding the effectiveness of these policies is important as policy makers and the society at large seek to achieve an optimal health outcome in the fight against the pandemic at the lowest economic cost.

We exploit between and within country variation in the type, timing, and level of intensity of lockdown policies in a multiple events study approach, which aims at disentangling the effect of each intervention on COVID-19 incidence and mobility patterns, while controlling for the presence of concurrent policy measures during the event window of the policy of interest, as well as for time fixed effects, day of week fixed effects, lagged COVID-19 prevalence, region fixed effects and time-varying country-specific characteristics. 

The main contributions of the paper are the following. First, we develop a multiple events model which allows us to identify the dynamic effects of each intervention while taking into account the presence of other concurrent interventions at each event time. Accounting for confounding policies is important because it allows us to avoid attributing the effect of other interventions to the policy of interest, and in addition to establish that it is policies that affect mobility patterns and not that policies ex-post respond to changing mobility patterns in the population.

Second, we consider a wide range of interventions across 135 countries, which vary in their type, intensity, and timing. The policy responses in focus are i) international travel controls, ii) public transport closures, iii) cancelation of public events, iv) restrictions on private gatherings, v) school closures, vi) workplace closures, vii) stay-at-home requirements and viii) internal mobility restrictions (across cities and regions). 

Third, we link policy interventions to mobility patterns by studying not only the impact of these policies on the incidence of COVID-19, but also on the time spent in a number of types of places such as i) retail and recreation, ii) grocery and pharmacy, iii) parks, iv) transit stations, v) the workplace and  vi) residential areas. Each of these types of places is characterized by different epidemiological features and, therefore, has a different potential for viral transmission. The mobility data can then also be viewed as a measure of compliance to the policies introduced and a mediator between policies and the spread of the disease.

The main result of the paper is that cancelling public events and imposing restrictions on private gatherings followed by school closures have quantitatively the most pronounced effects. They are followed by workplace as well as stay-at-home requirements, whose statistical significance and levels of effect are not as pronounced. Instead, we find no effects for international travel controls, public transport closures and restrictions on movements across cities and regions. We thus establish i) the order in which public policies help curb the pandemic and ii) that these effects are mediated by the way they change population mobility patterns in a manner consistent with time-use and epidemiological factors.

The rest of the paper is structured as follows. Section \ref{section:lit} contains a literature review, while Section \ref{section:data} discusses the data and presents summary statistics about NPIs and mobility patterns. Section \ref{section:model} describes the model and identification issues. The results are presented in Section \ref{section:results}, which contains a discussion linking the evidence on COVID-19 incidence with mobility patterns. Section \ref{section:conclusion} concludes by summarizing the findings and discussing future research.

\section{Literature} \label{section:lit}
\noindent
Research on infectious diseases focuses on vaccinations and drugs but it also aims at curbing the spread of the diseases by understanding and predicting their spatiotemporal dynamics, especially in the event of a new virus outbreak. Recent epidemiological models have been enriched to incorporate the impact of NPIs on these dynamics, which are at the core of this paper. The canonical model used in epidemiology is the so-called SIR model (\cite{doi:10.1098/rspa.1927.0118}). It provides a simple and relevant representation of the mechanics of virus propagation with three categories of individuals: i) the people who are susceptible to become infected (the S subpopulation), ii) the infected who can transmit the disease (the I) and iii) those who have recovered and cannot infect anymore (the R). A crucial concept in the SIR model is the $R0$, which is the average number of people that a sick person infects before she recovers. While the R0 is often considered as a biological characteristic of the virus’ transmissibility, it is also affected by environmental, behavioral, and social dimensions, including NPI’s.
 
In its basic form, the SIR compartmental model assumes that the population of interest is homogeneous in terms of exposure, immunity and chances of recovery. This assumption, however, is not realistic as in practice these factors have proven to be key in guiding policy interventions (\cite{auchincloss2012review}). Relaxing this assumption gave rise to extensions of this model which aim at capturing the multiple dimensions of heterogeneity by partitioning the population into groups based on age or location. Pushed to the extreme, such partitions lead to the individual-based models (\cite{eubank2004modelling}), which require data on the intensity of contacts between individuals of different age groups to calibrate person-to-person contact rates, for instance via social mixing matrices. Using this approach, \cite{jarvis} find for the UK that lockdown policies reduced the average number of daily contacts by $73$ percent, resulting in a drop of the $R0$ from $2.6$ to $0.62$, while  \cite{singh2020age} show for India that lockdown policies are unlikely to be effective if applied for $3$ weeks or less.

Beyond the partitioning of the population by age groups or communities, the compartmental model has been extended to take into account important specificities of the disease, such as the incubation period, the duration of the acquired immunity, or the challenges it presents given the current state of knowledge. In the context of the COVID-19 pandemic, several variants of the compartmental model have been used. The SEIR model takes into account the group of “exposed” individuals who can be asymptomatic carriers during the incubation period (\cite{karin2020adaptive}, \cite{pang2020public}; \cite{roy2020covid},  \cite{lyra2020covid}, \cite{lai2020effect}). In the SIRS model, recovery only provides a short-lived immunity, so that the R group moves back to the S group after some time (\cite{ ng2020covid19}). The SIOR model considers that only a fraction of the infected group is detected, or “observed”, by healthcare services (\cite{scala2020between}). While these various models capture different important features of the COVID-19 pandemic and provide predictions on the impact of NPI’s on the spread of the disease, their results are usually simulation-based and rely on structural assumptions.

In the field of economics, recent contributions depart from simple versions of the SIR model and introduce confinement policies as well as economic concepts (e.g. incentives, economic cost of lockdown, value of life). These papers highlight through calibrated simulations the tradeoffs between the mortality induced by the excess demand for healthcare services and the economic losses induced by confinement policies. \cite{Gonzalez} develop a SIR model in which policies take into account, among others, the rate of time preference, the learning of healthcare services and the severity of output losses. \cite{garibaldi} depart from the observation that the SIR model treats transitions from S to I as exogenous. In other words, the SIR model does not take into account individuals’ decision to reduce the intensity of their contacts and their exposure to the virus. The authors borrow concepts from the search and matching model (\cite{pissarides2000}) to introduce a contact function into the SIR model with forward-looking agents. They show that the decentralized epidemic equilibrium is likely to be suboptimal due to the presence of externalities: while individuals care about the private benefits of distancing, they neglect its social benefits and the fact that it reduces the risk of hospital congestion; on the other hand, from a dynamic perspective, they do not take into account the benefits of herd immunity. \cite{greenstone} develop a method to quantify the economic benefits of social distancing measures in terms of lives saved. They find that 1.7 million lives could be saved by applying mild social distancing for 3 to 4 months, which they estimate to be worth 8 trillion dollars accruing for 90 percent to the population above 50 years of age. \cite{barro2020non} studies the impact of NPI’s in the US during the Great Influenza Pandemic at the end of 1918 finding that even though NPIs reduced deaths peaks, and thereby reduced the stress imposed on healthcare services, they failed to significantly decrease overall mortality, which is likely due to the short application of the NPI’s, with an average duration of around one month.

The papers related to this study are \cite{chen}, \cite{gao2020mapping}, \cite{engle2020staying} and \cite{huber}. \cite{chen} focus on the reproduction number, which depends on the timing of NPI's with a parametric time lag effect, and predict for 9 countries the transmission dynamics under various sets of NPI's showing that country differences lead to different optimal policies with heterogeneous tradeoffs between health and economic costs. By combining geographic information systems and daily mobility patterns in US counties, derived from smartphone location big data, \cite{gao2020mapping} show that in many counties in which mobility restrictions were only recommended but not imposed, mobility did not decrease. \cite{engle2020staying} using US county-level GPS and COVID-19 cases,  study the impact of local disease prevalence and confinement orders on mobility solving a utility maximization problem after splitting the utility derived from traveling a unit of distance into costs independent from the epidemic and costs related to perceived risk of contracting the disease. They find substantial effects of local infection rates, while official confinement orders lead to a mobility reduction of less than 8 percent. \cite{huber} exploit regional variation in Germany and Switzerland to assess the impact of the timing of COVID-19 response measures finding that a relatively later exposure to the measures entails higher cumulative hospitalization and death rates.

We differ from these papers in the following ways: i) we develop a multiple events model exploiting the timing, type and level of intensity of several public policies with the advantage of flexibility in the non-parametric estimation of their dynamic impacts, taking into account the contemporaneous presence of multiple interventions; ii) we consider as outcomes both the COVID-19 cases and various mobility patterns, with the latter capturing how often and how long certain public places or one's residence is frequented; and iii) we analyze a panel dataset of $135$ countries. 

\section{Data and Descriptives}  \label{section:data}
\noindent
The analysis combines information from multiple data sources on (i) the non-pharmaceutical interventions implemented by governments, (ii) the daily number of infections, (iii) the evolution of population's mobility patterns, and (iv) various country characteristics.

Non-pharmaceutical interventions are collected by the Oxford COVID-19 Government Response Tracker (henceforth OxCGRT) for most countries of the world. The OxCGRT gathers publicly available information on several indicators of public policies aiming at mitigating the propagation of the virus. We focus on the following interventions: i) international travel controls, ii) closure of public transport, iii) cancelation of public events, iv) restrictions on private gatherings, v) closure of schools, vi) closure of workplaces, viii) restrictions on internal movement and viii) stay-at-home requirements. For each of these policies, we exploit information on the dates of introduction as well as qualitative time-varying information on their intensity. Intensity is measured in a scale from 1 to 6, which reflects whether the intervention is (i) recommended, (ii) mandatory with some flexibility, and (iii) mandatory with no flexibility, and whether it is geographically targeted or applied to the entire country. Recommended policies which are targeted obtain a value of 1, while mandatory policies with no flexibility applied to the entire country obtain a value of 6, with values in between referring to combinations of the policy stringency and its geographic scope.\footnote{To fix ideas, when the schooling policy receives an intensity score of 4, it means that it was not made mandatory in all schools or in all education levels, but it was applied to the entire country. A score of 5 means that it was made mandatory to all schools and education levels, but only in some areas of the country. A score of 6 means that it is mandatory for all schools and areas of the country. The average intensity level across countries is 2.9 for international travel controls, 3.8 for public transport closures, 5.4 for school closures, 3.8 for workplace closures, 4.7 for cancelling public events, 4 for restrictions on private gatherings, 3.2 for stay-at-home requirements and 4.2 for internal movement restrictions.} We use a sample of 135 countries for the estimation of NPIs, which is the set of countries for which we also have information on country characteristics (for a complete list see Appendix \ref{appendix:sample}).

Figure \ref{fig:days_policy_summary} presents the distribution of the number of days it took for each policy to be introduced after the first COVID-19 case averaged across countries. The distribution for the international travel controls is bimodal with the first mode well ahead of the first case.  All policies have a main mode close to zero, with cancelation of public events and school closures enacted earlier, followed by restrictions on private gatherings and workplace closures, stay-at-home requirements, internal mobility restrictions and public transportation restrictions, and a secondary mode late into the epidemic. 

The number of confirmed cases of COVID-19 infections is extracted from the ECDC, which examines reports from health authorities worldwide in a systematic way in order to produce the number of COVID-19 cases and deaths every day. This provides us with information on the number of new cases each day in each country. The observed variation in the incidence of COVID-19 cases may be influenced in part by variation in reporting. In order to remove such random variation from the data, we use a 3-day moving average of the confirmed new cases and the inverse hyperbolic sine transformation in order to include days with zero reported new cases.\footnote{Using the 3-day moving average helps visualization without changing the main findings. We also considered a 7-day moving average, which similarly maintains the main findings while removing reporting idiosyncrasies but additionally flatten features of the data which might be of interest. We opted for the 3-day moving average as the middle ground. We also conduct our analysis without the 3-day moving average as discussed below.}

To study how mobility patterns have evolved worldwide, we resort to Google's Community Mobility Reports. The Google mobility data are created with {\it ``aggregated, anonymized sets of data from users who have turned on the Location History setting''} on their phone and show how {\it ``visits and length of stay''} at different types of places change compared to the median value, for the corresponding day of the week, during the 5-week period from January 3, 2020 to February 6, 2020.\footnote{\href{https://www.google.com/covid19/mobility/}{COVID-19 Community Mobility Reports.}} Google's ability to accurately locate phones and to correctly categorize places varies both across countries as well as within (urban vs. rural areas). These data contain information on various epidemiologically relevant categories of places such as: i) retail and recreation, ii) grocery and pharmacy, iii) parks, iv) workplaces, v) transit stations and vi) residential areas. Retail and recreation covers visits to restaurants, cafes, shopping centers, theme parks, museums, libraries, and movie theaters. Grocery and pharmacy covers grocery markets, food warehouses, farmers markets, specialty food shops, drug stores, and pharmacies. Parks encompass national parks, public beaches, marinas, dog parks, plazas, and public gardens. Transit stations cover subway, bus and train stations. From the sample of 135 countries we have information on mobility patterns from Google for a subsample of 108 countries (for a complete list see Appendix \ref{appendix:sample}).

Figure \ref{fig:mobility_summary} presents the distribution of these data averaged across countries both before and after the first confirmed COVID-19 case. Before the first case, all distributions are highly concentrated around zero, which suggests no substantial change of movement compared to the baseline period. After the first case, retail and recreation as well as transit stations have a mean of just above -40, suggesting a 40 percentage points drop, while differing in their variance. Grocery and pharmacy as well as workplaces have a mean of around -20 differing in their skewness (grocery and pharmacy is heavy on the left). Parks stand out for having a mean closest to zero and being heavy on the right. Finally, residential areas have a mean of just over 10.\footnote{Considering that staying at home is by far the most time intensive activity, according to time-use studies, this value is quite large. We expand on this point in Section \ref{results:consolidated}.} In terms of densities, retail and recreation, grocery and pharmacy, transit stations as well as workplaces are somewhat similar, while parks and residential areas are on their own on opposite sides.

\section{The model}\label{section:model}
\noindent
We follow an event-study approach around the time of policy implementation, which we extend to account for multiple events. The single event study (e.g. \cite{kleven2019children}) can be expressed with the following equation: 
\begin{equation}\label{eq:single}
\begin{split}
Y_{c,t}= & \displaystyle{\sum_{j \neq-20}} \alpha_{j}  I[j=t-t_c^{\pi}]
+\displaystyle{\sum_{l}} \gamma_l I[l=t] + \\
& \displaystyle{\sum_{d}}\delta_d I[d_c(t) = d]+
\displaystyle{\sum_{r}}\rho_r R_r + 
\phi Z_{c,t-1} + \theta X_{c}  + 
\epsilon_{c,t},
\end{split}
\end{equation}
\noindent
where $Y_{c,t}$ denotes the outcome in country $c$ at event time $t$. The first term, on the right hand side, is a set of event time dummies for the intervention of interest $\pi$, where $t_c^{\pi}$ denotes its implementation day in country $c$. We consider the outcome in the window starting 20 days before the intervention up to 35 days after its implementation, so the event time runs from $-20$ to $+35$. We omit the event time dummy at $j=-20$ so that the event time coefficients of interest $\alpha_{j}$ measure the impact of intervention $\pi$ at time $j$ relative to the twenty days before the policy was implemented. 

The second term, on the right hand side of equation (\ref{eq:single}), is a set of dummies which control non-parametrically for trends in the time since the first-observed COVID-19 case. Identifying the coefficients of the event time dummies conditional on these time effects is possible because the timing of NPIs differs across countries. The third term, is a set of day-of-week dummies controlling for potential day-specific differences both in terms of reporting of new cases and of mobility patterns (here $d_c(t)$ returns the day of the week for event time $t$ in country $c$). The fourth term, is a set of dummies for the following regions: Europe, Asia, Middle East, North America, South America, Oceania and Africa. The fifth term, is the log value of the total number of confirmed cases at $t-1$ in country $c$. Including this variable allows us to capture the size of the pool of infected people, which is a crucial factor both when the outcome is the incidence of new cases, in line with the SIR framework, as well as when it is mobility patterns, as populations may react to the perceived threat of contamination.\footnote{Adjusting the total number of confirmed cases by the number of deaths does not affect our main results.} The sixth term, is a set of country specific variables controlling for differences across countries, such as per capita GDP, population density and the urbanization rate, followed by the error term.\footnote{When the outcome is the incidence of COVID-19, these controls are epidemiologically relevant, whereas, when we consider Google mobility types as our outcome they help control for differences in Google's ability to geo-locate phones and detect types of places.}

When evaluating the effect of the intervention of interest $\pi$, it is important to take into account the presence of other contemporaneous interventions, which can have their own contribution in affecting the outcome, and thus, if ignored can lead to biased estimates. However, identifying the effect of the policy of interest $\pi$ with multiple events is more challenging than in the single-event case, especially when the multiple events fully overlap during the event window of the policy of interest. Concurrent NPIs, denoted by $\pi'$, can be controlled for by introducing in equation (\ref{eq:single}) a new term, $F^{\pi'}[j=t-t_c^{\pi}] $, which is a set of dummies - one dummy for each event time of the policy of interest $\pi$ -  which are equal to one if any other interventions $\pi'$ are in effect at event time $j$ for country $c$. The multiple events regression equation can then be written as follows: 
\begin{equation}\label{eq:multiple}
\begin{split}
Y_{c,t} = & \displaystyle{\sum_{j \neq-20}} \alpha_{j}  I[j=t-t_c^{\pi}] + 
\displaystyle{\sum_{j}}\beta_j  F^{\pi'}[j=t-t_c^{\pi}] +
\displaystyle{\sum_{l}} \gamma_l I[l=t] + \\
& \displaystyle{\sum_{d}}\delta_d I[d_c(t) = d]+
\displaystyle{\sum_{r}}\rho_r R_r + 
\phi Z_{c,t-1} +
 \theta X_{c}  + 
\epsilon_{c,t}.
\end{split}
\end{equation}

The identification problem in the multiple-event case emerges as soon as other policies have been introduced before the start of event window of policy ${\pi}$. This would result in a complete overlap of policies within the event window, making it impossible to separately identify the effect of the event of interest from the other contemporaneous events.\footnote{When other policies are enacted within the event window, then the two set of event dummies are not perfectly collinear so the coefficient estimates $\alpha_j$ and $\beta_j$ can be separately identified, but at the cost of high variance because of multicollinearity.}  

To achieve identification in the multiple events case, we use the level of intensity of each policy which varies both within and across policies, as well as across countries. This variation of policy intensity allows us to identify separately the effect of the policy of interest $\pi$, while taking into account other concurrent NPIs, $\pi'$. The extended multiple events regression equation can be written as follows:
\begin{equation}\label{eq:imultiple}
\begin{split}
Y_{c,t}= & \displaystyle{\sum_{j \neq-20}} \alpha_{j}  S^{\pi}[j=t-t_c^{\pi}]  + 
\displaystyle{\sum_{j}}\beta_j  \bar{S}^{\pi'}[j=t-t_c^{\pi}]  +
\displaystyle{\sum_{l}} \gamma_l I[l=t] + \\
&  \displaystyle{\sum_{d}}\delta_d I[d_c(t) = d]+
\displaystyle{\sum_{r}}\rho_r R_r + 
\phi Z_{c,t-1} +
 \theta X_{c}  + 
\epsilon_{c,t},
\end{split}
\end{equation}
\noindent
where the first term, $S^{\pi}[j=t-t_c^{\pi}] $, is taking the value of the level of intensity of the policy of interest $\pi$ in country $c$ at event time $j$ and zero otherwise, while the second term, $\bar{S}^{\pi'}[j=t-t_c^{\pi}] $, is equal to the average level of intensity of all other contemporaneous policies $\pi'$ of country $c$ at the event time $j$, and zero if there are no other policies active at that event time. That is, we extend equation (\ref{eq:multiple})  in two ways: 1) we multiply the event dummies for policy $\pi$ with the intensity level of the policy at event time $j$ - in other words,  $I[j=t-t_c^{\pi}] $  in equation (\ref{eq:multiple}) generalizes to $S^{\pi}[j=t-t_c^{\pi}] $ in equation (\ref{eq:imultiple}); and 2) we multiply the dummies controlling for the presence of any other policies $\pi'$ - at event time $j$ for policy $\pi$ - with their average intensity at event time $j$ - in other words,  $F^{\pi'}$ in equation (\ref{eq:multiple}) generalizes to $\bar{S}^{\pi'}_{c,t} $ in equation (\ref{eq:imultiple}). 

Our identification relies on the variation in the timing and intensity of various interventions both within and across countries, conditional on the prevalence of COVID-19, time effects since the first observed case, day effects, country-specific characteristics and continent effects. This variation allows to separately identify the effect of intervention $\pi$ from that of other concurrent NPIs, $\pi'$. The coefficient estimates $a_j$ in equation \ref{eq:imultiple} measure the unit level intensity effect of policy $\pi$ at event time $j$ on the outcome.

\section{Results} \label{section:results}
\noindent
This section contains the results split in three subsections. Subsection \ref{results:cases} contains results on the effect of NPIs on the incidence of COVID-19, whereas Subsection \ref{results:mobility} contains results on the effect of NPIs on population mobility patterns from Google's Community Mobility Reports. Finally, Subsection \ref{results:consolidated} provides a consolidated view on the link between the impact of NPIs on new cases through their effect on mobility.

\subsection{Lockdown Policies on COVID-19 Incidence }\label{results:cases}
 \noindent
We start by comparing the estimates for the dynamic effects of each intervention obtained from the two versions of the model: i) ignoring concurrent interventions, i.e. estimating equation (\ref{eq:imultiple}) without the second term, and ii) controlling for concurrent interventions, i.e. estimating  equation (\ref{eq:imultiple}).\footnote{We focus on the results where the dependent variable is the 3-day moving average of confirmed new cases. The estimates with the number of confirmed cases are reported in Figures \ref{fig:cases_1_i_single_ss_1_cc_1} and \ref{fig:cases_1_i_multiple_ss_1_cc_1} in Appendix \ref{appendix:figures}.}

\begin{figure}[!htb]
	\centering
 \includegraphics[width=.75\textwidth]{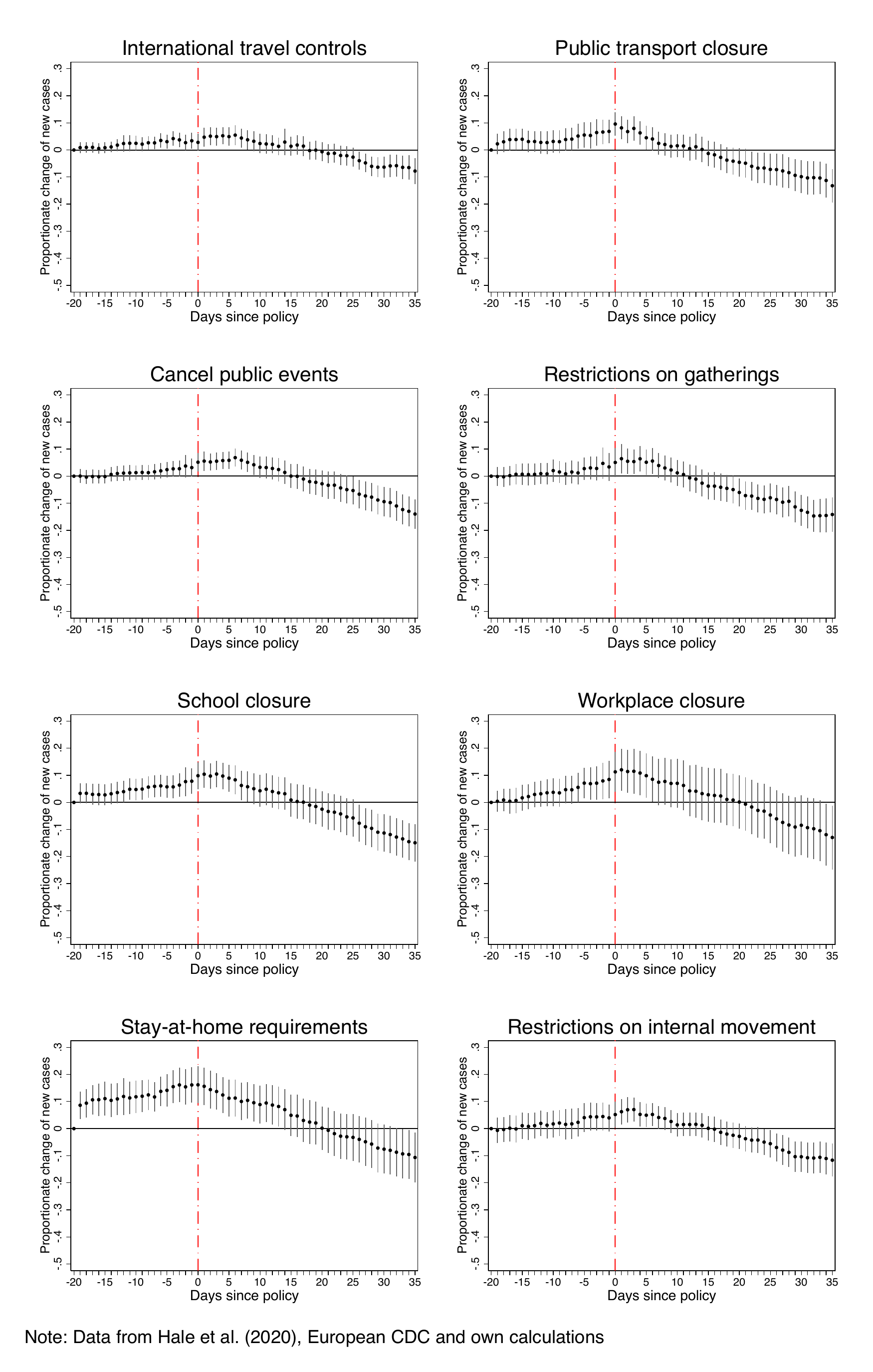}
	\caption{Effects of lockdown policies on {\bf COVID-19} confirmed new cases (3-day moving average, in logs) without concurrent policy controls.}
	\label{fig:macases_1_i_single_ss_1_cc_1}
\end{figure}

Comparing the two sets of estimates, it becomes apparent that ignoring the presence of other interventions leads to biased estimates. Specifically, the results of Figure \ref{fig:macases_1_i_single_ss_1_cc_1},  which report the estimates ignoring concurrent policies, suggest that all policies tend to have a significant impact following a similar pattern. That is, in the days preceding the introduction of the policy, the incidence of COVID-19 increases until it reaches a peak after few a days following its introduction. Then, the number of new cases per day start to decrease, and within a month they become significantly lower than the reference event time (20 days before the policy). 

\begin{figure}[!htb]
	\centering
  \includegraphics[width=.75\textwidth]{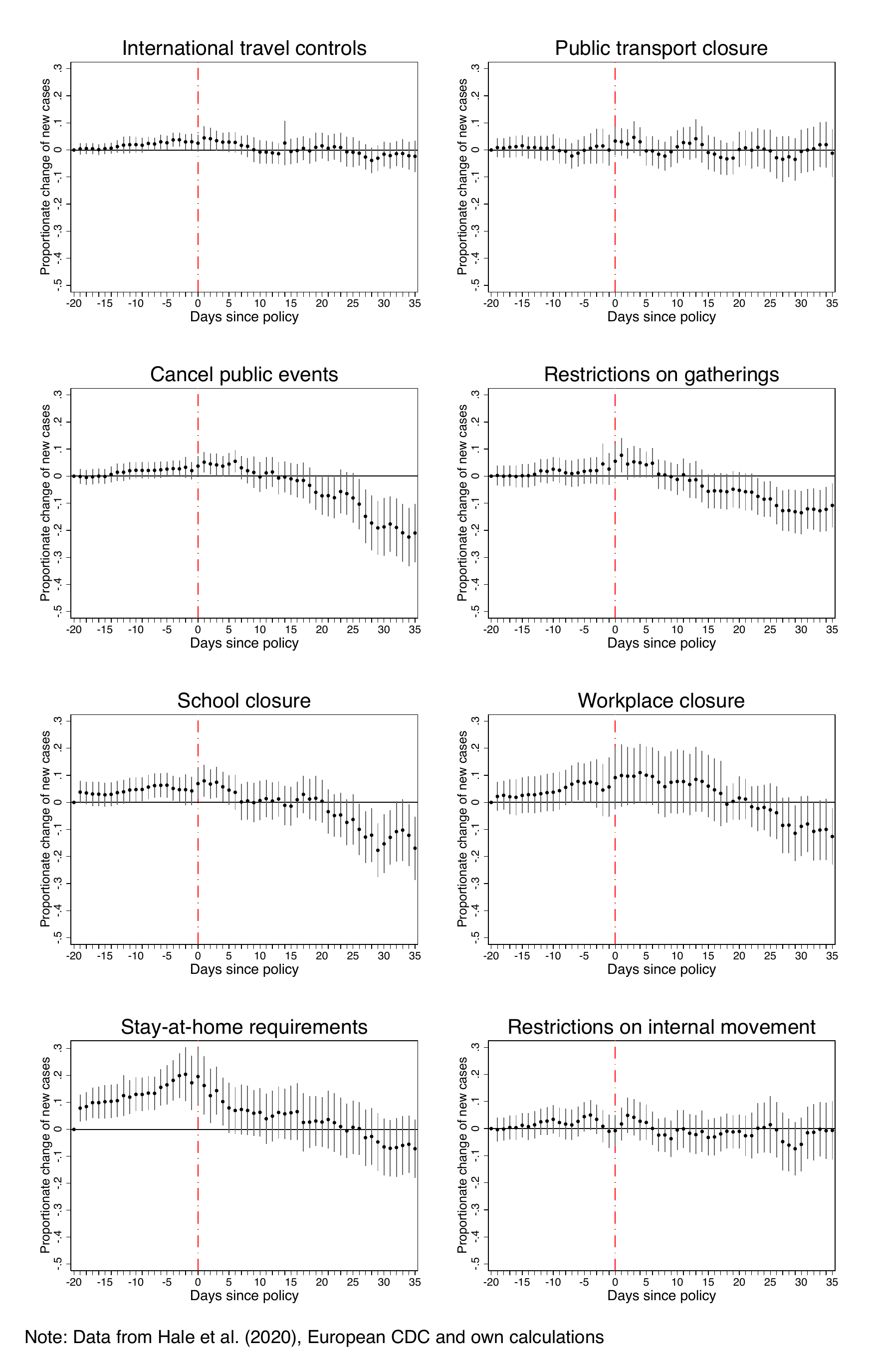}
	\caption{Effects of lockdown policies on {\bf COVID-19} confirmed new cases (3-day moving average, in logs) with controls for concurrent policies.}
	\label{fig:macases_1_i_multiple_ss_1_cc_1}
\end{figure}

The analysis without controlling for other concurrent policies seems to suggest that all interventions were successful in containing new infections. However, the estimates in Figure \ref{fig:macases_1_i_multiple_ss_1_cc_1}, which are obtained after controlling for concurrent policies, convey a different message. This is especially true for the two transport related interventions, i.e. \textit{international travel controls} and \textit{public transport closure}, and for \textit{restrictions on internal movement}, which have almost no impact on new cases once other interventions are controlled for.

The two policies with the largest effects, which are robust to confounding by other policies, are \textit{cancelling of public events} and \textit{restrictions on private gatherings}. These are policies which aim to reduce massive contacts.\footnote{See the discussion in Section \ref{results:consolidated}.} For both, we observe a drop in the incidence of COVID-19 starting about one week after implementation, which becomes significantly different than zero within two weeks. Around the end of the event window, a unit increase in the intensity of the policy of interest leads to a 20\% decrease in the number of new infections in the case of public events cancelation, and a decrease of about 12\% in the case of restrictions on private gatherings, compared to the reference event time.

\textit{School} and \textit{workplace} closures aim to control contacts in large groups, but unlike public events and private gatherings are easier to monitor and regulate as well as trace whenever infections do occur. We find that new infections start declining a few days after school closures, with the effect becoming negative and significant about 25 days after implementation. Around the end of the event window, a unit increase in the intensity of school closures leads to about a 15\% drop of new infections compared to the reference event time. For workplace closures, we find that new infections start declining starting from the second week after implementation and the effect becomes negative and significant only towards the end of the event window, with a unit increase in the policy intensity leading to about a 10\% drop of new infections.

Finally, \textit{stay-at-home requirements} aim to impose mobility constraints at the individual level, which is arguably the most extreme of all measures and was generally introduced when infections were reaching alarming growth rates. This is captured in Figure \ref{fig:macases_1_i_multiple_ss_1_cc_1} which shows that, around the date of introduction of the policy, there were on average 20\% more new cases every day compared to the reference event time, with the policy reversing that trend immediately. By the end of the window, the coefficient estimates are statistically significantly lower from those around the policy implementation day, although they are not statistically different from zero.

\subsection{Lockdown Policies on Google Mobility Patterns}\label{results:mobility}
\begin{figure}[!htb]
	\centering
 \includegraphics[width=.87\textwidth]{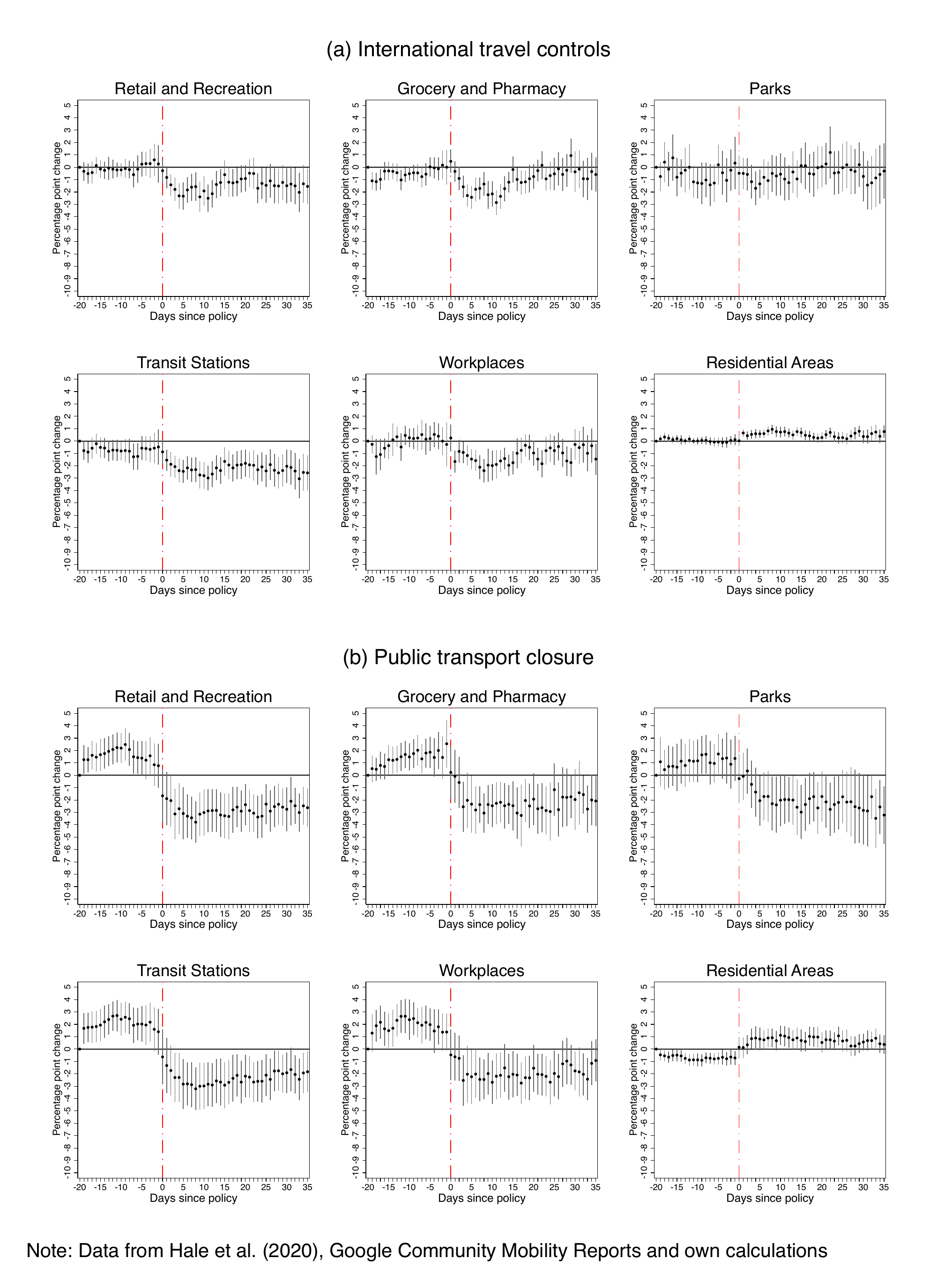}
	\caption{Effects of {\bf international travel controls} (panel a) and {\bf closure of public transportation} (panel b) on Google mobility patterns. }
	\label{fig:travel_and_transport_i_multiple_ss_1_cc_1}
\end{figure}

\begin{figure}[!htb]
	\centering
 \includegraphics[width=.87\textwidth]{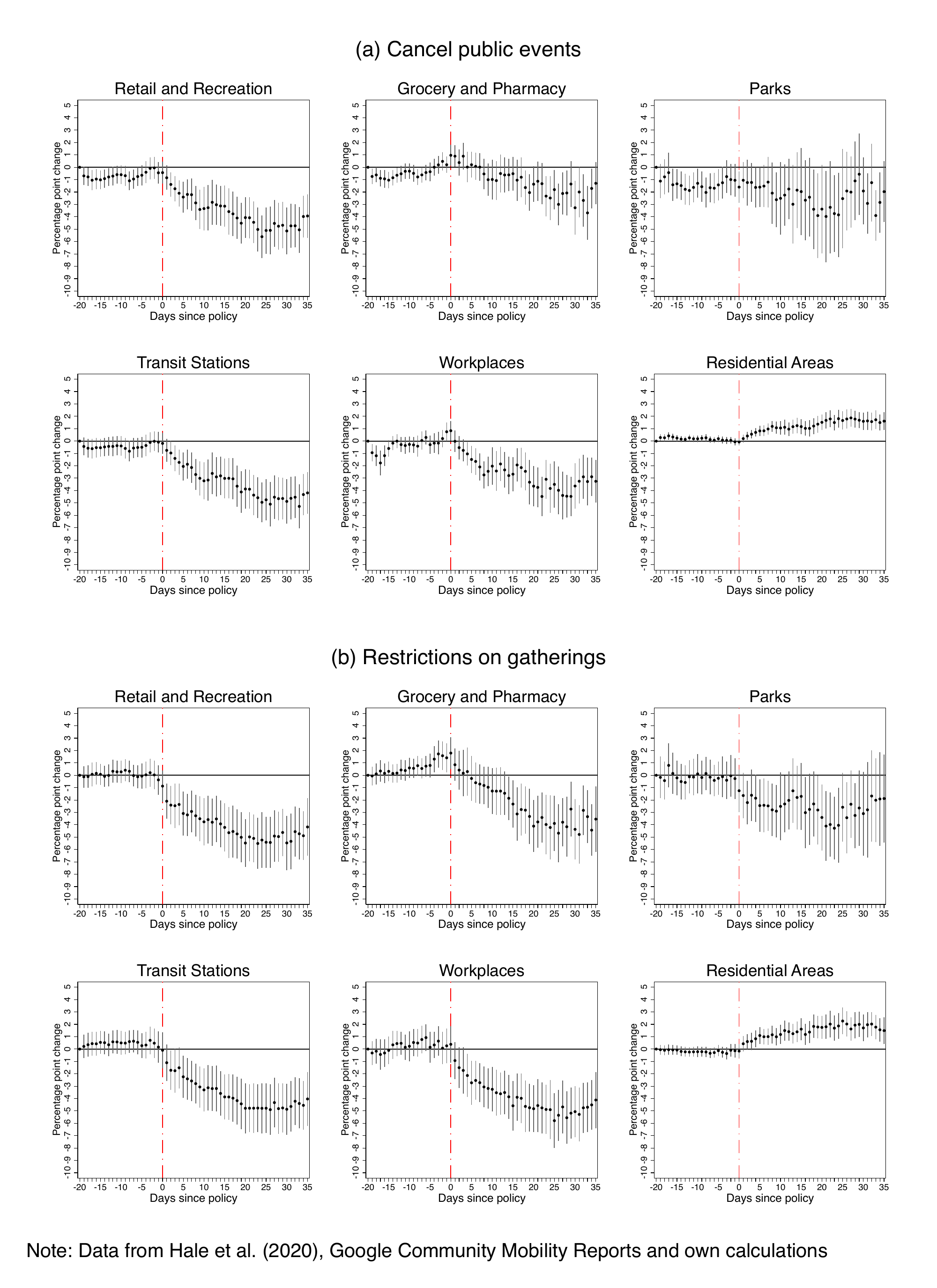}
	\caption{Effects of {\bf public events cancellations}  (panel a) and {\bf  restrictions on gatherings} (panel b) on Google mobility patterns. }
	\label{fig:events_and_gatherings_i_multiple_ss_1_cc_1}
\end{figure}

\begin{figure}[!htb]
	\centering
 \includegraphics[width=.87\textwidth]{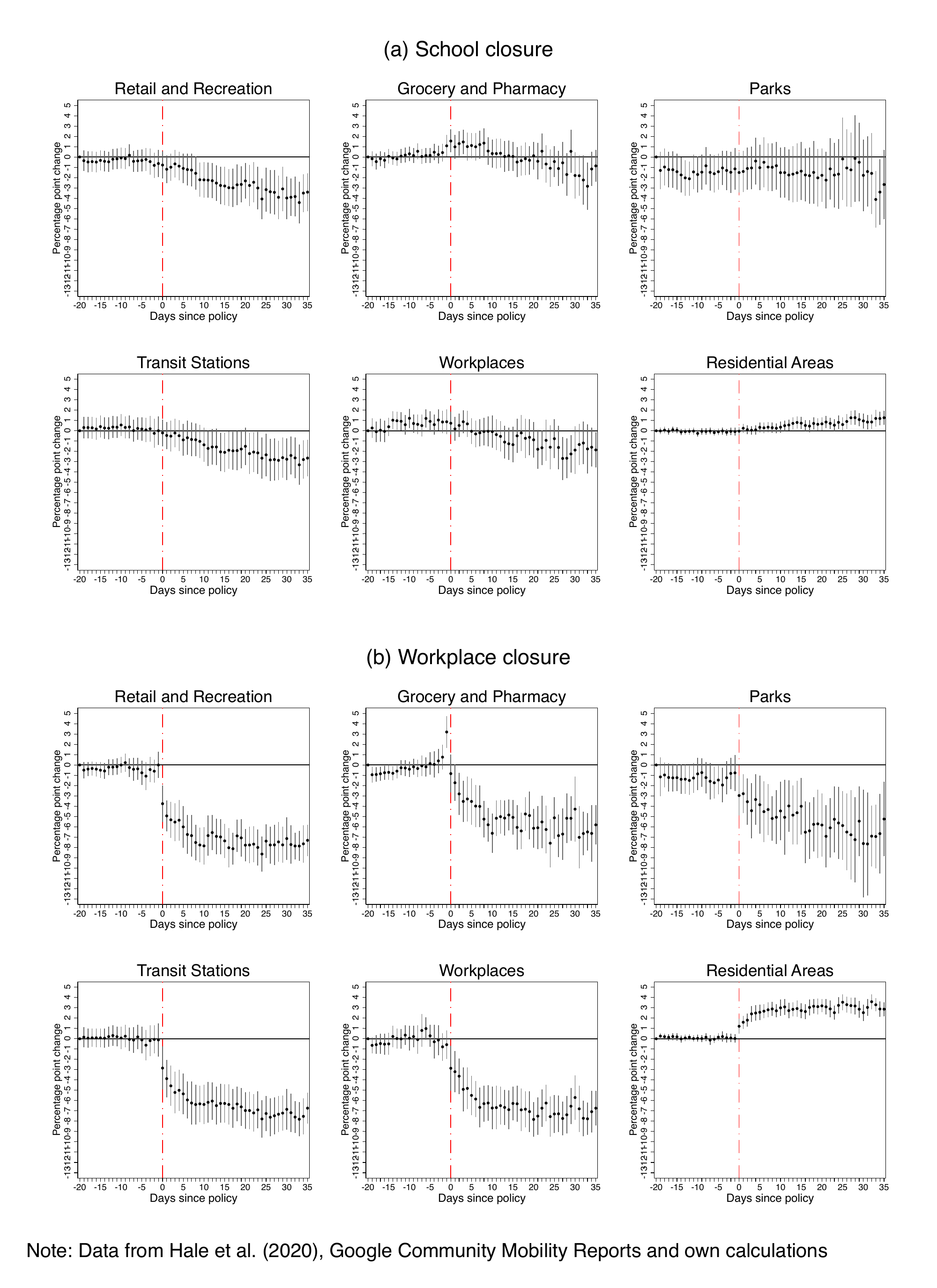}
	\caption{Effects of  {\bf school}  (panel a) and {\bf workplace} (panel b) closures on Google mobility patterns. }
	\label{fig:school_and_workplace_i_multiple_ss_1_cc_1}
\end{figure}

\begin{figure}[!htb]
	\centering
 \includegraphics[width=.87\textwidth]{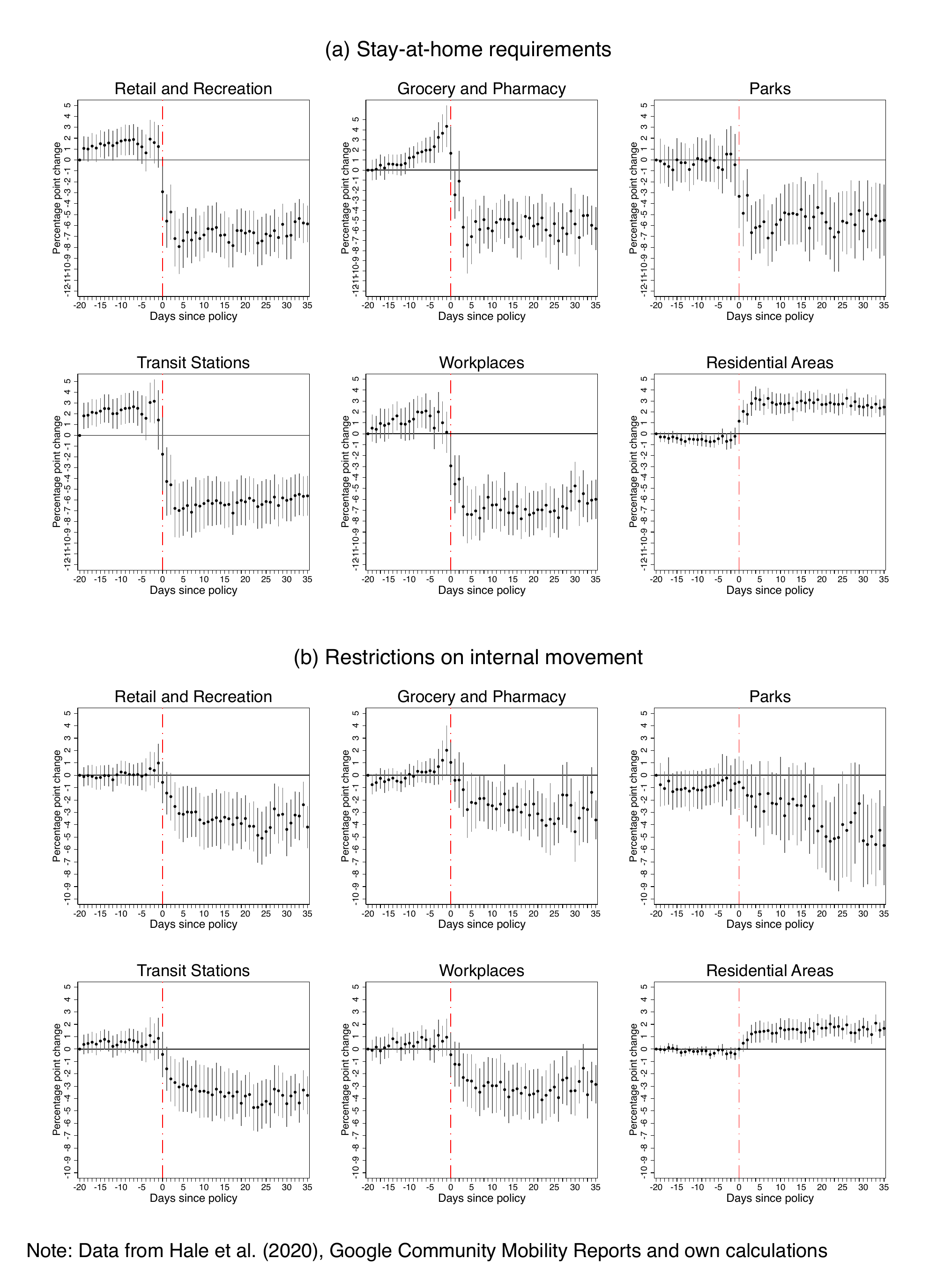}
	\caption{Effects of {\bf stay-at-home requirements}  (panel a) and {\bf restrictions on internal mobility} (panel b) on Google mobility patterns. }
	\label{fig:home_and_mobility_i_multiple_ss_1_cc_1}
\end{figure}

\begin{figure}[!htb]
	\centering
 \includegraphics[width=.85\textwidth]{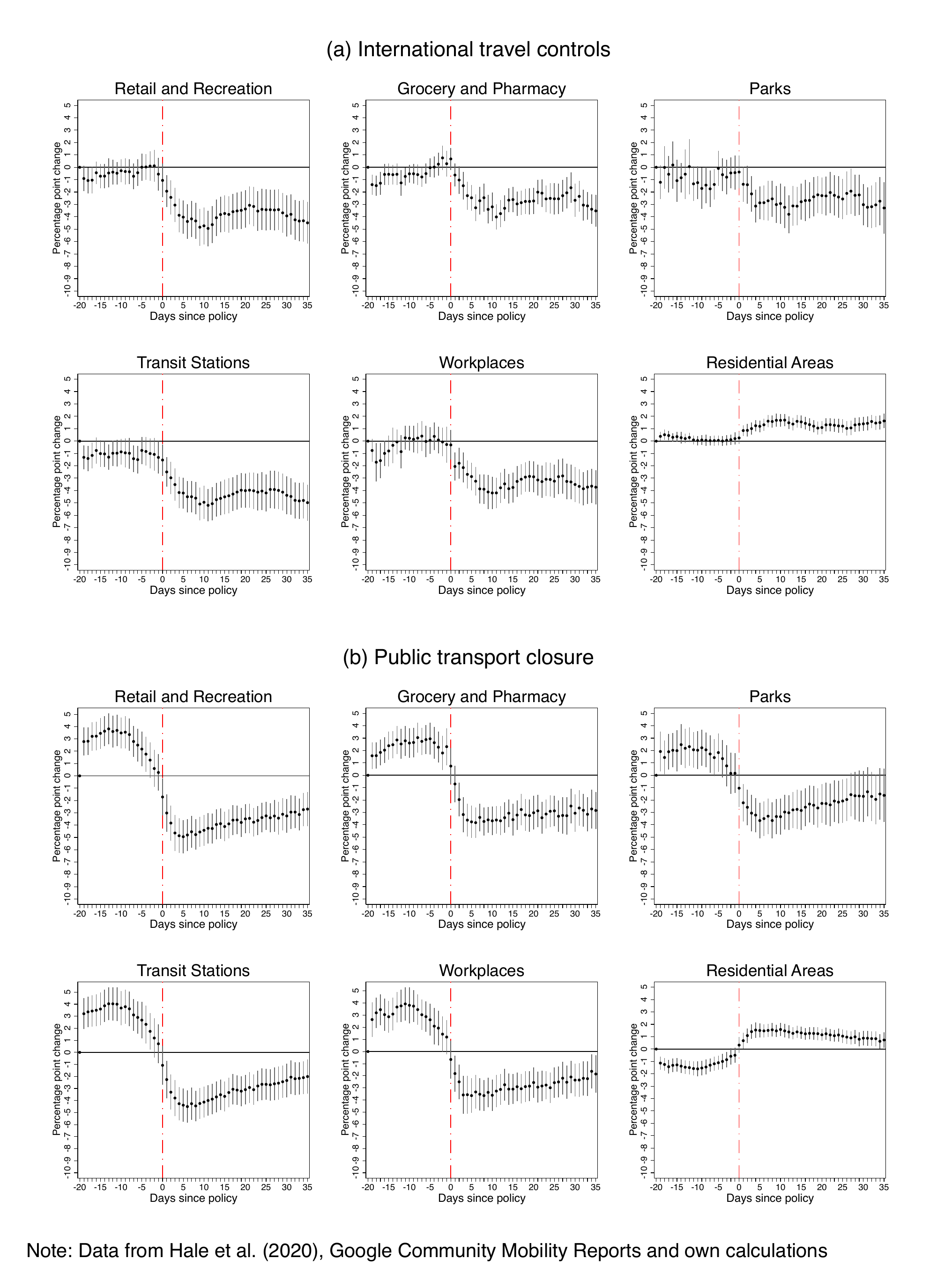}
	\caption{Effects of {\bf international travel controls} (panel a) and {\bf public transportation closure} (panel b) on Google mobility patterns without concurrent policy controls. }
	\label{fig:travel_and_transport_i_single_ss_1_cc_1}
\end{figure}

\begin{figure}[!htb]
	\centering
 \includegraphics[width=.85\textwidth]{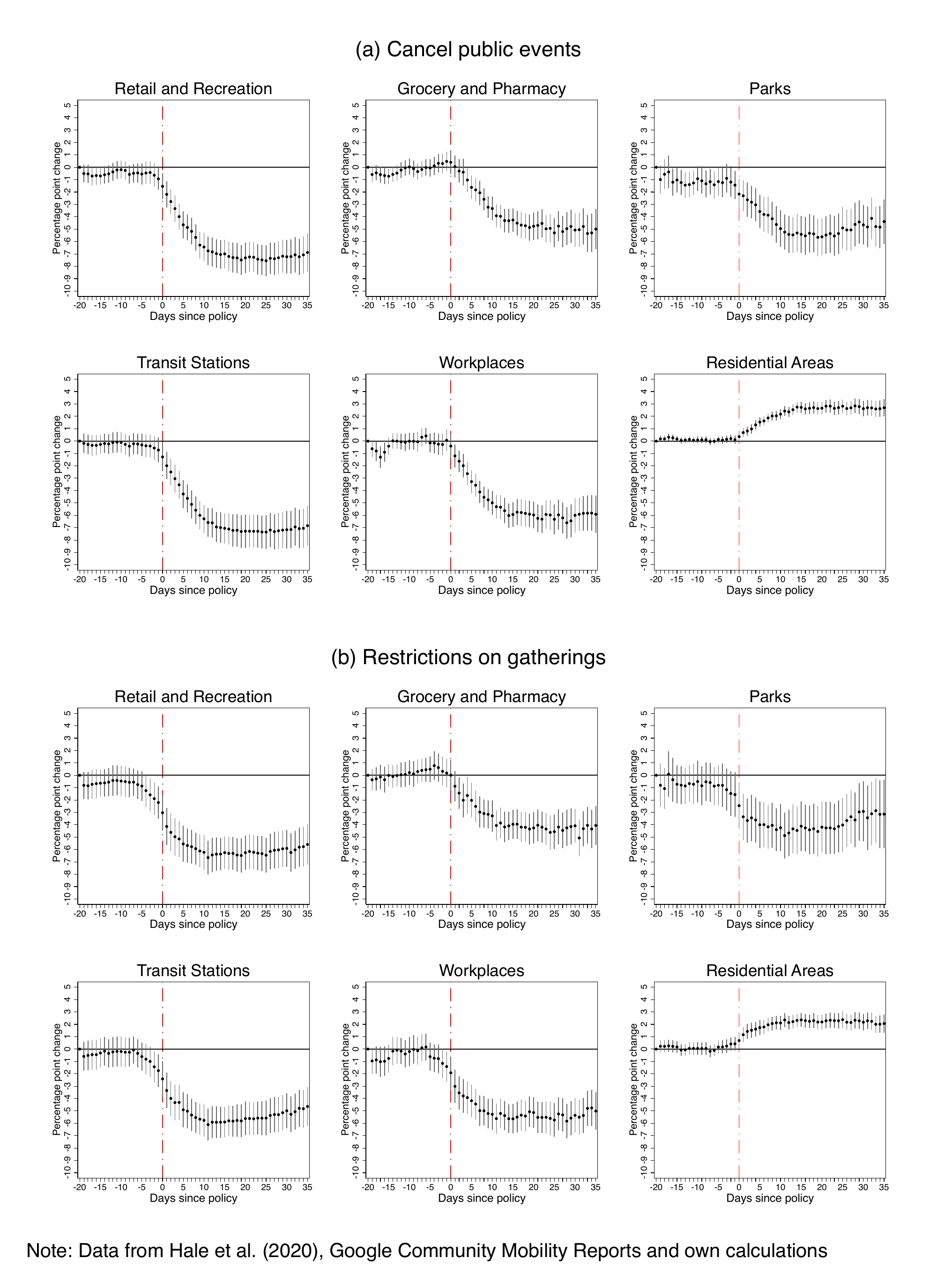}
	\caption{The effect of {\bf public events cancellations} (panel a) and {\bf  restrictions on gatherings} (panel b) on Google mobility patterns without concurrent policy controls. }
	\label{fig:events_and_gatherings_i_single_ss_1_cc_1}
\end{figure}

\noindent
Google mobility patterns are observed as percentage point deviations from a reference calendar period before the onset of COVID-19. As a result, the coefficient estimates of interest - first term of equation  (\ref{eq:imultiple}) - measure the percentage point change in mobility patterns for a unit level of intensity of each intervention compared to the reference point before implementation. Similar to COVID-19 confirmed new cases, we obtain estimates both with, as well as without controls for other ongoing interventions. The estimates with controls for concurrent policies are presented in Figures \ref{fig:travel_and_transport_i_multiple_ss_1_cc_1}, \ref{fig:events_and_gatherings_i_multiple_ss_1_cc_1}, \ref{fig:school_and_workplace_i_multiple_ss_1_cc_1} and \ref{fig:home_and_mobility_i_multiple_ss_1_cc_1}, while those without controlling for confounding policies can be found in Figures \ref{fig:travel_and_transport_i_single_ss_1_cc_1}, \ref{fig:events_and_gatherings_i_single_ss_1_cc_1}, \ref{fig:school_and_workplace_i_single_ss_1_cc_1} and \ref{fig:home_and_mobility_i_single_ss_1_cc_1} in Appendix \ref{appendix:figures}.

We find that, when we do not control for confounding policies, right after the day of policy implementation there is a general pattern of sharp and large drops in all mobility patterns related to activities undertaken outside residential areas, and an increase in the amount of time spent in the place of residence. However, once we control for other concurrent NPIs, many of these effects are either much smaller, or sometimes not significantly different from zero.  For example, the estimates for \textit{international travel controls} without accounting for confounders, shown in panel (a) of Figure \ref{fig:travel_and_transport_i_single_ss_1_cc_1}, suggest a significant decline in movements immediately after the policy implementation across most places (retail and recreation, grocery and pharmacy, parks, transit stations, workplaces) and increases in staying home. However, after controlling for other interventions present around the same time, we find in panel (a) of Figure \ref{fig:travel_and_transport_i_multiple_ss_1_cc_1} that restrictions in international travel have a much smaller impact on all types of movement. 

After accounting for multiple events, panel (b) of Figure \ref{fig:travel_and_transport_i_multiple_ss_1_cc_1} shows that restrictions on \textit{public transportation} lead to a sharp discontinuity at the day of the intervention with lower movements outside home. Interestingly, the strength of this decrease in mobility tends to weaken with time. The limited - and not very persistent - reductions in mobility patterns observed for international travel controls and closure of public transportation are consistent with the small effects of these policies on the incidence of new cases reported in Section \ref{results:cases}.

The \textit{cancelation of public events} and \textit{restrictions on private gatherings}, which led to the most important reductions in new infections, as reported in Section \ref{results:cases}, also exhibit large and persistent negative impacts on retail and recreation, transit stations, workplaces and to a lesser extent grocery and pharmacy (panels (a) and (b) of Figure \ref{fig:events_and_gatherings_i_multiple_ss_1_cc_1}). For both policies, the magnitude of these drops is around 5 percentage points per unit level of policy intensity. These findings are consistent with the fact that attending public events and private gatherings generate spillover effects on various activities outside the homeplace. Conversely, these policies have significantly increased time spent at home. 

As reported in Subsection \ref{results:cases}, the set of interventions with the second strongest reductions on subsequent infections were  \textit{school} and \textit{workplace closures}. Figure \ref{fig:school_and_workplace_i_multiple_ss_1_cc_1}  shows that these policies do change the mobility trends associated with crowded places, such as retail and recreation, transit stations and workplaces. Again, this can be explained by the fact that closing schools and workplaces generate spillover effects on other activities. The sharpness and the magnitude of the mobility decreases is much stronger in workplace than in school closures, with a unit level increase in the intensity of the policy leading to a stable decline of up to about 7-8 and 2-3 percentage points, respectively. This difference is also consistent with the following observations. First, workers generally have access to more mobility patterns and activities than pupils. Second, while pupils staying at home might constrain the mobility of one parent, closing workplaces affects the mobility of all adults working in the household.

\textit{Stay-at-home requirements} (panel (a) of Figure \ref{fig:home_and_mobility_i_multiple_ss_1_cc_1}) result in large drops in all population mobility patterns at the time when they were introduced, a fact which is consistent with the reversal of the increasing trend of new cases reported in Subsection \ref{results:cases}. Finally, in line with the results obtained for infection cases, \textit{internal mobility restrictions} (panel (b) of Figure \ref{fig:home_and_mobility_i_multiple_ss_1_cc_1}) have a similar impact on all mobility patterns as home confinement, though neither as sharp, nor as strong. The magnitude of these effects are about half the size compared to stay-at-home requirements. 

We conclude this section with three remarks. First, mobility patterns do not exhibit much in the way of anticipation effects once we control for confounding NPIs. This suggests that it is policies affecting mobility patterns and not that policies ex-post responding to de facto changing mobility patterns in the population. It is worth noting that estimates ignoring concurrent NPIs would have led to a completely different conclusion; as shown in Figures \ref{fig:travel_and_transport_i_single_ss_1_cc_1} to \ref{fig:home_and_mobility_i_single_ss_1_cc_1} in Appendix \ref{appendix:figures}, for several interventions mobility patterns seem to respond before policies are in place. Second, we observe a spike in movements to groceries and pharmacies prior to the introductions of several NPIs, such as public transport and workplace closure, as well as stay-at-home and internal movement restrictions. This is consistent with the widely reported runs on the shelves in anticipation of lockdowns, concerns about imminent shortages, as well as with inadvertent signaling from these interventions about the threat level of the pandemic. Again, we are able to detect these mobility patterns only when we account for confounding policies. In light of the fact that we control for the state of the epidemic by using lags of total confirmed cases, this result is robust and shows the strength of our model.  Third, it appears that the decline in mobility patterns is stable towards the last days of analysis, suggesting that compliance does not decline over time, at least within the 35-day window of our study.

\subsection{Lockdown policies: a consolidated view}\label{results:consolidated}
\noindent
In this section, we expand on how the observed variation in the effects of NPIs on the incidence of COVID-19, reported in Section \ref{results:cases}, can be understood by the way in which they affect various mobility patterns across places, reported in Section \ref{results:mobility}, which differ in a number of characteristics as they pertain to epidemiology, as well as in their time-use intensity. We thus provide a consistent framework for our results. 

First, the degree to which restricting mobility to different places is expected to affect new infections depends on several characteristics of these places, where the most important are the following: i) numerosity, ii) density, iii) social norms, iv) geographical range and v) tracking ability. For example, more numerous and dense places, such as large private gatherings and public events, are more likely to contribute to new infections because the two-meter safe social-distancing rule is more likely to be violated there than say in parks. However, places with similar density can be conducive to different behavior types due to social norms; for example, in a soccer game, where there are large numbers of people densely brought together, there are different norms of accepted behavior compared to the regulated environment of a workplace. Furthermore, places such as schools vs. transit stations, or public events, can have different epidemiological range. For example, an infection at school has a range of perhaps a couple of kilometers (students reside close to their schools), while in the case of a soccer game it might be several kilometers and even cross country borders. Finally, places differ in terms of how easy it is to trace an infection back whenever it occurs, which is important because tracking contains the spread of the virus. For example, an incident at a workplace can be announced immediately to employees and an ad hoc lockdown can be probably enforced at the same time, while an infection which occurs at a transit station is impossible to trace back or treat with a local lockdown.  
  
Second, places differ from a time-use perspective. Based on time-use surveys on how people spend their time in everyday life, for example, European adults in selected countries between the ages of 20 and 74 years old, spend on average on a daily basis: i) 15 hours at home preparing meals, sleeping, and on household activities, ii) a little less than 3 hours at work, which has mostly a large workplace component, iii) a little more than 1 hour traveling and commuting and iv) about 4 and a half hours on other activities including leisure (recreation, parks, home) and shopping (retail, groceries and pharmacies).\footnote{\href{https://ec.europa.eu/eurostat/documents/3930297/5953614/KS-58-04-998-EN.PDF}{https://ec.europa.eu/eurostat/documents/3930297/5953614/KS-58-04-998-EN.PDF}} These differences in time-use suggest, for example, that observing a 3 percentage points increase in time spent at the place of residence implies an increase of about half an hour, whereas a decrease of 8 percentage points in workplaces amounts to a drop of about 15 minutes. 

In light of these differences across places, we find that NPIs tend to reduce activities away from home, while increasing time spent at home to a varying degree depending on their time-use footprint, while their impact on the epidemic depends on the above mentioned epidemiologically relevant characteristics. 

More specifically, \textit{cancellation of public events}, and to a lesser extent \textit{restrictions on private gatherings}, which are seen to lead to a large reduction in new infections, are interventions that reduce exposure to numerous and dense locations, where contact tracing is difficult, and can have a large epidemiological range within and across countries. Similarly, \textit{stay-at-home requirements}, \textit{workplace} and \textit{school closures} reduce activities away from home and lead to significant reductions in the incidence of new infections, which nevertheless are not as large as for public events and private gatherings, possibly because of the differences in numerosity, density and ability to trace new infections in these environments. 

On the other hand, although \textit{restrictions on internal movements} reduce mobility across cities and regions, they impact the spread of the disease in a less pronounced way. This is consistent with the fact that these restrictions are not clearly linked to places with high density, and their potential to slow down new infections by restricting geographical mobility is reduced, once other policies such as workplace closures and restrictions on private gatherings are in place. Furthermore, \textit{public transport closures} were introduced on average at a time where demand for traveling and commuting has declined due to other restrictions in place such as workplace closures, which can explain both their limited impact on mobility and on reducing new infected cases. 

Finally, the limited impact of \textit{international travel controls}, although they were imposed relatively early by many countries, is likely explained by the lack of stringency of the controls. If countries have banned all international travel soon after the outbreak in China, it would have certainly be an effective measure to seal the country from the virus. However, because most countries did not introduce such bans before the virus has started spreading domestically, or they did introduce some restrictions but not complete bans, those restrictions had a limited impact on mobility and could only reduce new imported infections but not contain the spread of the virus.

\section{Conclusion} \label{section:conclusion}
\noindent
The COVID-19 pandemic impacts societies and economies in multiple and dramatic ways. The exact extent of this impact in economic and social terms is certainly going to remain a topic of interest in the time ahead. In this paper, we develop a multiple events model to study the effect of lockdown policies on the incidence of new infections and on mobility patterns exploiting variation in the type, timing and intensity of confinement policies across 135 countries. The key contributions of the paper are twofold: i) we model the dynamic effects of each policy on the incidence of new infections accounting for concurrent policies, while in line with the standard SIR model, we specify future infections (incidence) as a function of past cases (prevalence), as well as a number of risk related characteristics, such as GDP per capita, population, population density and urbanization rates, all of which enrich the exposure to risk of infection with heterogeneity within and between countries and ii) we link the effect of NPIs on new infections through their impact on mobility patterns.

Our findings establish that cancelling public events and enforcing restrictions on private gatherings followed by school closures, which reduce mobility patterns in numerous and dense locations, each with their own particular behavioral norms, have the largest effect on curbing the pandemic in terms of statistical significance and levels of effect. They are followed by workplace and stay-at-home requirements, which also reduce activities away from home and lead to significant reductions in the incidence of COVID-19, which nevertheless are not as large as for public events, private gatherings and school closures, possibly because of the differences in numerosity, density and the ability to trace new infections in these environments. Instead, restrictions on internal movement, public transport closures and international travel controls do not lead to a significant reduction of new infections. The limited impact of travel controls, although imposed relatively early in many countries, is likely explained by their lack of stringency allowing the virus to cross borders. 

Our econometric framework is suitable for the study of dynamic effects with multiple events, which can be applied in many settings. A natural one is the upcoming exit strategies from the lockdowns, which we will turn to next.
\bibliographystyle{apalike}
\bibliography{covid19-BKN.bib}

\begin{thebibliography}{23}
\providecommand{\natexlab}[1]{#1}
\providecommand{\url}[1]{\texttt{#1}}
\expandafter\ifx\csname urlstyle\endcsname\relax
  \providecommand{\doi}[1]{doi: #1}\else
  \providecommand{\doi}{doi: \begingroup \urlstyle{rm}\Url}\fi

\bibitem[Auchincloss et~al.(2012)Auchincloss, Gebreab, Mair, and
  Diez~Roux]{auchincloss2012review}
Amy~H Auchincloss, Samson~Y Gebreab, Christina Mair, and Ana~V Diez~Roux.
\newblock A review of spatial methods in epidemiology, 2000--2010.
\newblock \emph{Annual review of public health}, 33:\penalty0 107--122, 2012.

\bibitem[Barro(2020)]{barro2020non}
Robert~J Barro.
\newblock Non-pharmaceutical interventions and mortality in us cities during
  the great influenza pandemic, 1918-1919.
\newblock Technical report, National Bureau of Economic Research, 2020.

\bibitem[Chen and Qiu(2020)]{chen}
Xiaohui Chen and Ziyi Qiu.
\newblock Scenario analysis of non-pharmaceutical interventions on global
  covid-19 transmissions.
\newblock \emph{arXiv preprint arXiv:2004.04529}, 2020.

\bibitem[Engle et~al.(2020)Engle, Stromme, and Zhou]{engle2020staying}
Samuel Engle, John Stromme, and Anson Zhou.
\newblock Staying at home: mobility effects of covid-19.
\newblock \emph{COVID Economics}, \penalty0 (2020-4), 2020.

\bibitem[Eubank et~al.(2004)Eubank, Guclu, Kumar, Marathe, Srinivasan,
  Toroczkai, and Wang]{eubank2004modelling}
Stephen Eubank, Hasan Guclu, VS~Anil Kumar, Madhav~V Marathe, Aravind
  Srinivasan, Zoltan Toroczkai, and Nan Wang.
\newblock Modelling disease outbreaks in realistic urban social networks.
\newblock \emph{Nature}, 429\penalty0 (6988):\penalty0 180--184, 2004.

\bibitem[Gao et~al.(2020)Gao, Rao, Kang, Liang, and Kruse]{gao2020mapping}
Song Gao, Jinmeng Rao, Yuhao Kang, Yunlei Liang, and Jake Kruse.
\newblock Mapping county-level mobility pattern changes in the united states in
  response to covid-19.
\newblock \emph{Available at SSRN 3570145}, 2020.

\bibitem[Garibaldi et~al.(2020)Garibaldi, Moen, and Pissarides]{garibaldi}
Pietro Garibaldi, Espen~R Moen, and Christopher~A Pissarides.
\newblock Modelling contacts and transitions in the sir epidemics model.
\newblock \emph{Covid Economics Vetted and Real-Time Papers, CEPR}, 2020.

\bibitem[Gonzalez-Eiras and Niepelt(2020)]{Gonzalez}
M~Gonzalez-Eiras and D.~Niepelt.
\newblock On the optimal "lockdown" during an epidemic.
\newblock \emph{Covid Economics Vetted and Real-Time Papers, CEPR}, 2020.

\bibitem[Greenstone and Nigam(2020)]{greenstone}
Michael Greenstone and Vishan Nigam.
\newblock Does social distancing matter?
\newblock \emph{University of Chicago, Becker Friedman Institute for Economics
  Working Paper}, \penalty0 (2020-26), 2020.

\bibitem[Hale et~al.(2020)Hale, Webster, Petherick, Phillips, and
  Kira]{hale2020oxford}
Thomas Hale, Samuel Webster, Anna Petherick, Toby Phillips, and Beatriz Kira.
\newblock Oxford covid-19 government response tracker.
\newblock \emph{Blavatnik School of Government}, 2020.

\bibitem[Huber and Langen(2020)]{huber}
Martin Huber and Henrika Langen.
\newblock The impact of response measures on covid-19-related hospitalization
  and death rates in germany and switzerland, 2020.

\bibitem[Jarvis et~al.(2020)Jarvis, Van~Zandvoort, Gimma, Prem, Klepac, Rubin,
  Edmunds, working group, et~al.]{jarvis}
Christopher~I Jarvis, Kevin Van~Zandvoort, Amy Gimma, Kiesha Prem, Petra
  Klepac, G~James Rubin, W~John Edmunds, CMMID COVID-19 working group, et~al.
\newblock Quantifying the impact of physical distance measures on the
  transmission of covid-19 in the uk.
\newblock \emph{BMC Med}, 18\penalty0 (124), 2020.

\bibitem[Karin et~al.(2020)Karin, Bar-On, Milo, Katzir, Mayo, Korem, Dudovich,
  Yashiv, Zehavi, Davidovich, et~al.]{karin2020adaptive}
Omer Karin, Yinon~M Bar-On, Tomer Milo, Itay Katzir, Avi Mayo, Yael Korem, Boaz
  Dudovich, Eran Yashiv, Amos~J Zehavi, Nadav Davidovich, et~al.
\newblock Adaptive cyclic exit strategies from lockdown to suppress covid-19
  and allow economic activity.
\newblock \emph{medRxiv}, 2020.

\bibitem[Kermack et~al.(1927)Kermack, McKendrick, and
  Walker]{doi:10.1098/rspa.1927.0118}
William~Ogilvy Kermack, A.~G. McKendrick, and Gilbert~Thomas Walker.
\newblock A contribution to the mathematical theory of epidemics.
\newblock \emph{Proceedings of the Royal Society of London. Series A,
  Containing Papers of a Mathematical and Physical Character}, 115\penalty0
  (772):\penalty0 700--721, 1927.
\newblock \doi{10.1098/rspa.1927.0118}.
\newblock URL
  \url{https://royalsocietypublishing.org/doi/abs/10.1098/rspa.1927.0118}.

\bibitem[Kleven et~al.(2019)Kleven, Landais, and
  S{\o}gaard]{kleven2019children}
Henrik Kleven, Camille Landais, and Jakob~Egholt S{\o}gaard.
\newblock Children and gender inequality: Evidence from denmark.
\newblock \emph{American Economic Journal: Applied Economics}, 11\penalty0
  (4):\penalty0 181--209, 2019.

\bibitem[Lai et~al.(2020)Lai, Ruktanonchai, Zhou, Prosper, Luo, Floyd,
  Wesolowski, Zhang, Du, Yu, et~al.]{lai2020effect}
Shengjie Lai, Nick~W Ruktanonchai, Liangcai Zhou, Olivia Prosper, Wei Luo,
  Jessica~R Floyd, Amy Wesolowski, Chi Zhang, Xiangjun Du, Hongjie Yu, et~al.
\newblock Effect of non-pharmaceutical interventions for containing the
  covid-19 outbreak: an observational and modelling study.
\newblock \emph{medRxiv}, 2020.

\bibitem[Lyra et~al.(2020)Lyra, do~Nascimento, Belkhiria, de~Almeida, Chrispim,
  and de~Andrade]{lyra2020covid}
Wladimir Lyra, Jose~Dias do~Nascimento, Jaber Belkhiria, Leandro de~Almeida,
  Pedro~Paulo Chrispim, and Ion de~Andrade.
\newblock Covid-19 pandemics modeling with seir (+ caqh), social distancing,
  and age stratification. the effect of vertical confinement and release in
  brazil.
\newblock \emph{medRxiv}, 2020.

\bibitem[Ng and Gui(2020)]{ng2020covid19}
Kok~Yew Ng and Meei~Mei Gui.
\newblock Covid-19: Development of a robust mathematical model and simulation
  package with consideration for ageing population and time delay for control
  action and resusceptibility, 2020.

\bibitem[Pang(2020)]{pang2020public}
Weijie Pang.
\newblock Public health policy: Covid-19 epidemic and seir model with
  asymptomatic viral carriers, 2020.

\bibitem[Pissarides(2000)]{pissarides2000}
Christopher~A Pissarides.
\newblock \emph{Equilibrium unemployment theory}.
\newblock MIT press, 2000.

\bibitem[Roy(2020)]{roy2020covid}
Shovonlal Roy.
\newblock Covid-19 pandemic: Impact of lockdown, contact and non-contact
  transmissions on infection dynamics.
\newblock \emph{medRxiv}, 2020.

\bibitem[Scala et~al.(2020)Scala, Flori, Spelta, Brugnoli, Cinelli,
  Quattrociocchi, and Pammolli]{scala2020between}
Antonio Scala, Andrea Flori, Alessandro Spelta, Emanuele Brugnoli, Matteo
  Cinelli, Walter Quattrociocchi, and Fabio Pammolli.
\newblock Between geography and demography: Key interdependencies and exit
  mechanisms for covid-19.
\newblock \emph{Available at SSRN 3572141}, 2020.

\bibitem[Singh and Adhikari(2020)]{singh2020age}
Rajesh Singh and R~Adhikari.
\newblock Age-structured impact of social distancing on the covid-19 epidemic
  in india.
\newblock \emph{arXiv preprint arXiv:2003.12055}, 2020.

\end{thebibliography}

\newpage

\section*{\bf  \center Figures} \label{section:figures}
\begin{figure}[!htb]
	\centering
  \includegraphics[width=.685\textwidth]{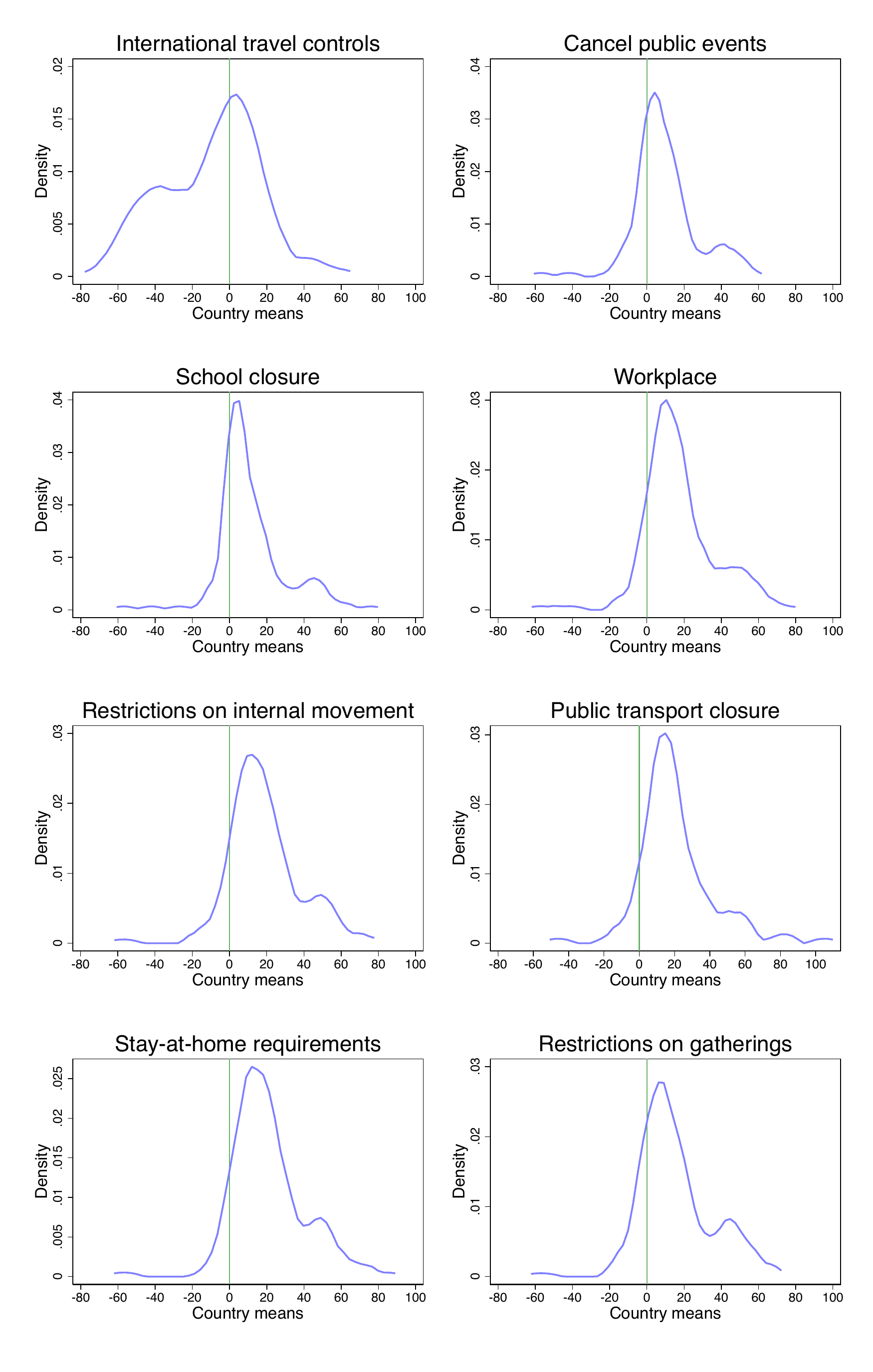}
	\caption{ Lockdown Policies - Days after first COVID-19 case each policy was introduced.}
	\label{fig:days_policy_summary}
\end{figure}

\newpage
\begin{figure}[!htb]
	\centering
  \includegraphics[width=\textwidth]{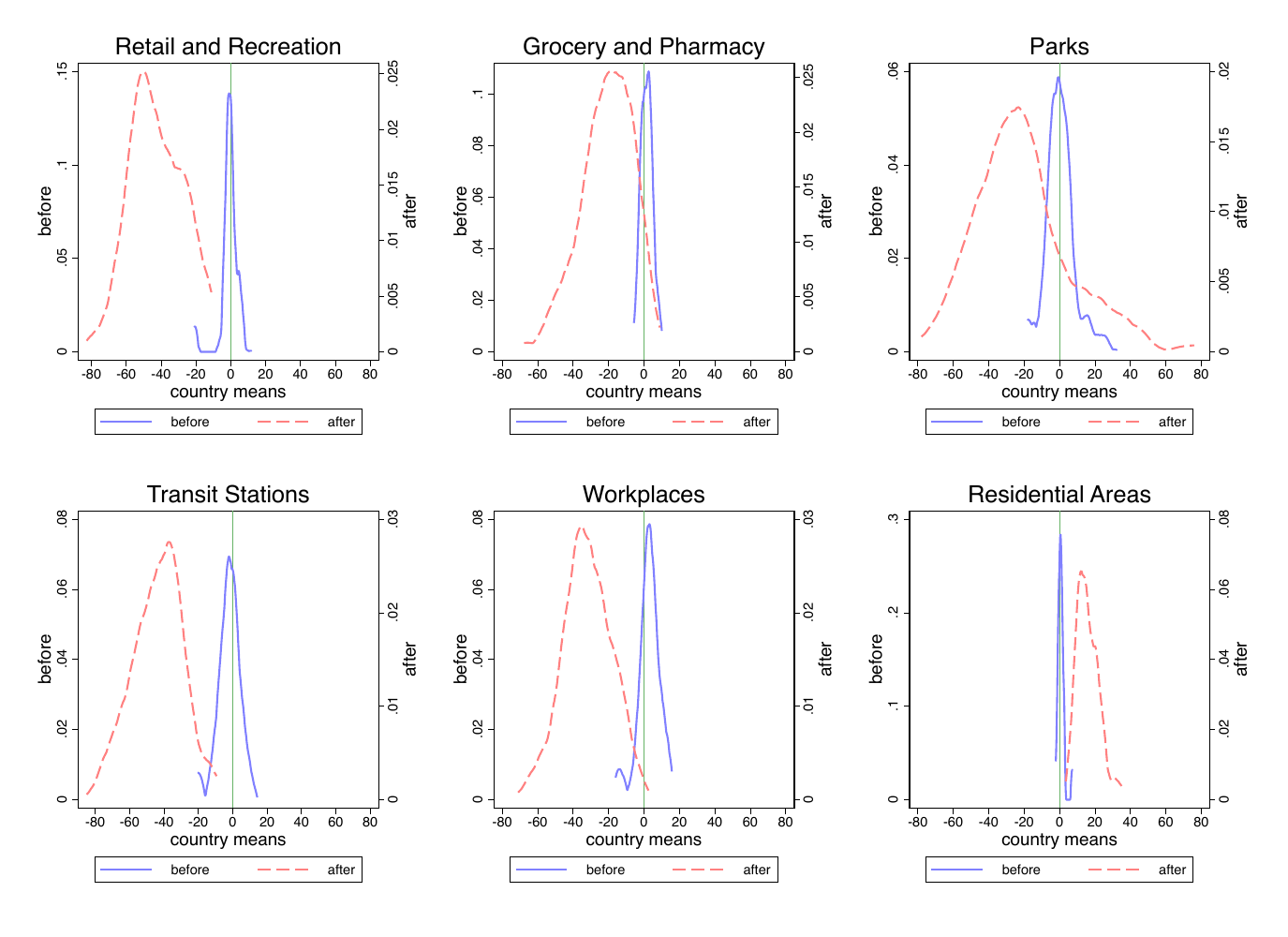}
	\caption{Google Mobility Patterns - Densities before and after first COVID-19 case.}
	\label{fig:mobility_summary}
\end{figure}


\newpage

\newpage

\newpage



\newpage


\newpage



\newpage


\appendix
\counterwithout{figure}{section}
\counterwithin{figure}{section}
\newpage
\section{Appendix - Figures} \label{appendix:figures}

\begin{figure}[!htb]
	\centering
  \includegraphics[width=.77\textwidth]{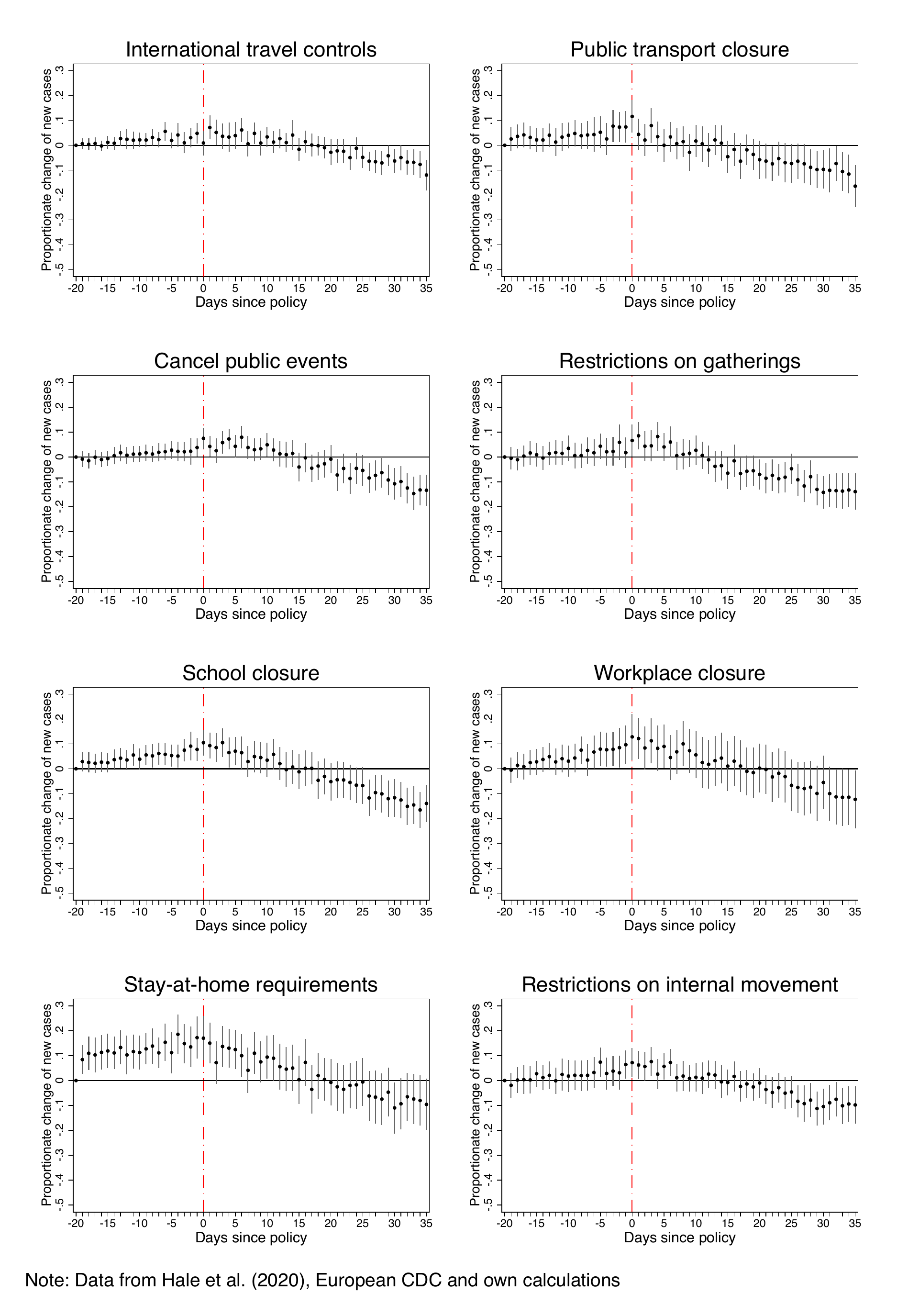}
	\caption{Effects of lockdown policies on {\bf COVID-19} confirmed new cases (in logs) without controlling for concurrent policies.}
	\label{fig:cases_1_i_single_ss_1_cc_1}
\end{figure}

\begin{figure}[!htb]
	\centering
  \includegraphics[width=.77\textwidth]{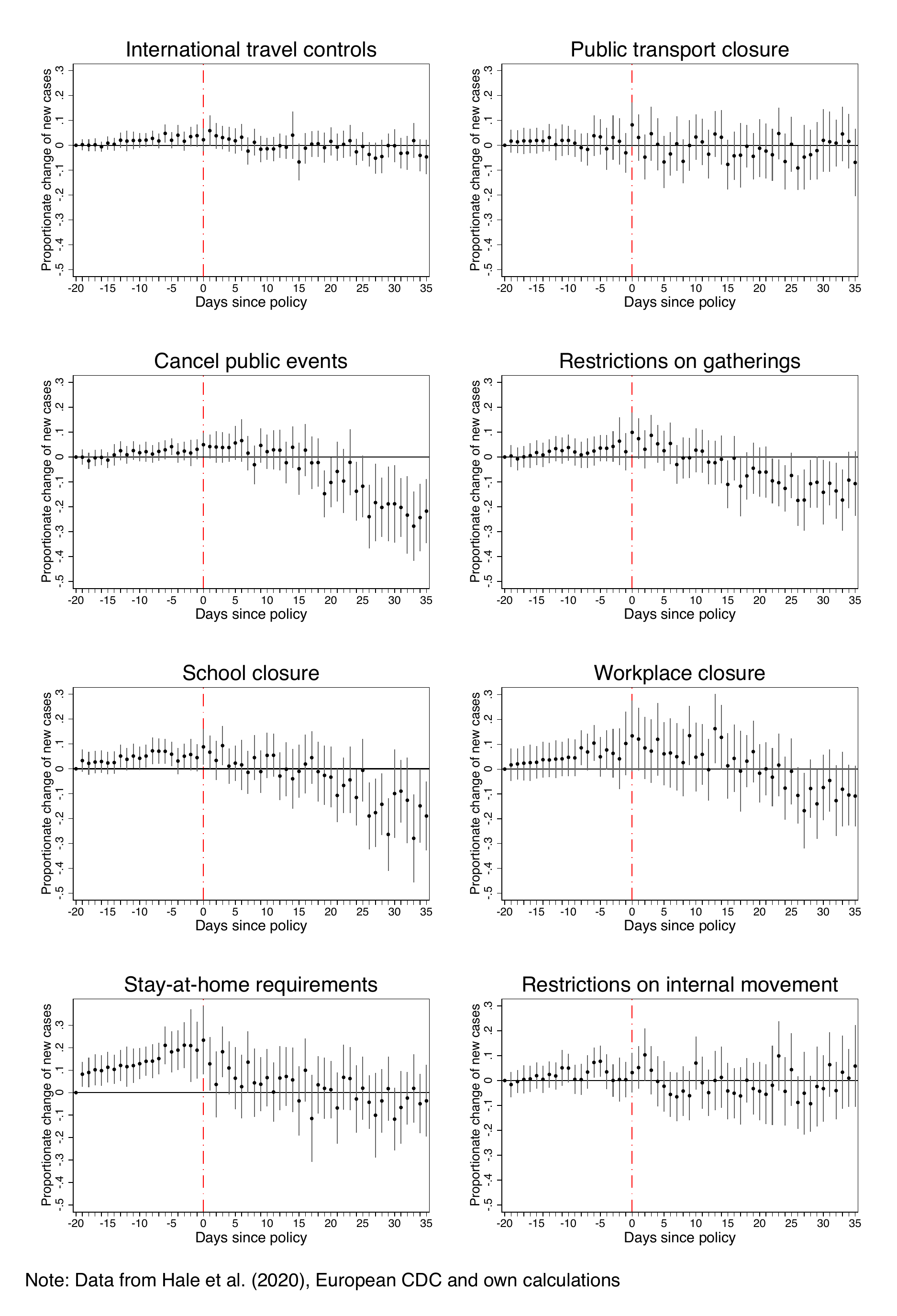}
	\caption{Effects of lockdown policies on {\bf COVID-19} confirmed new cases (in logs) controlling for concurrent policies.}
	\label{fig:cases_1_i_multiple_ss_1_cc_1}
\end{figure}



\newpage


\newpage



\begin{figure}[!htb]
	\centering
 \includegraphics[width=.85\textwidth]{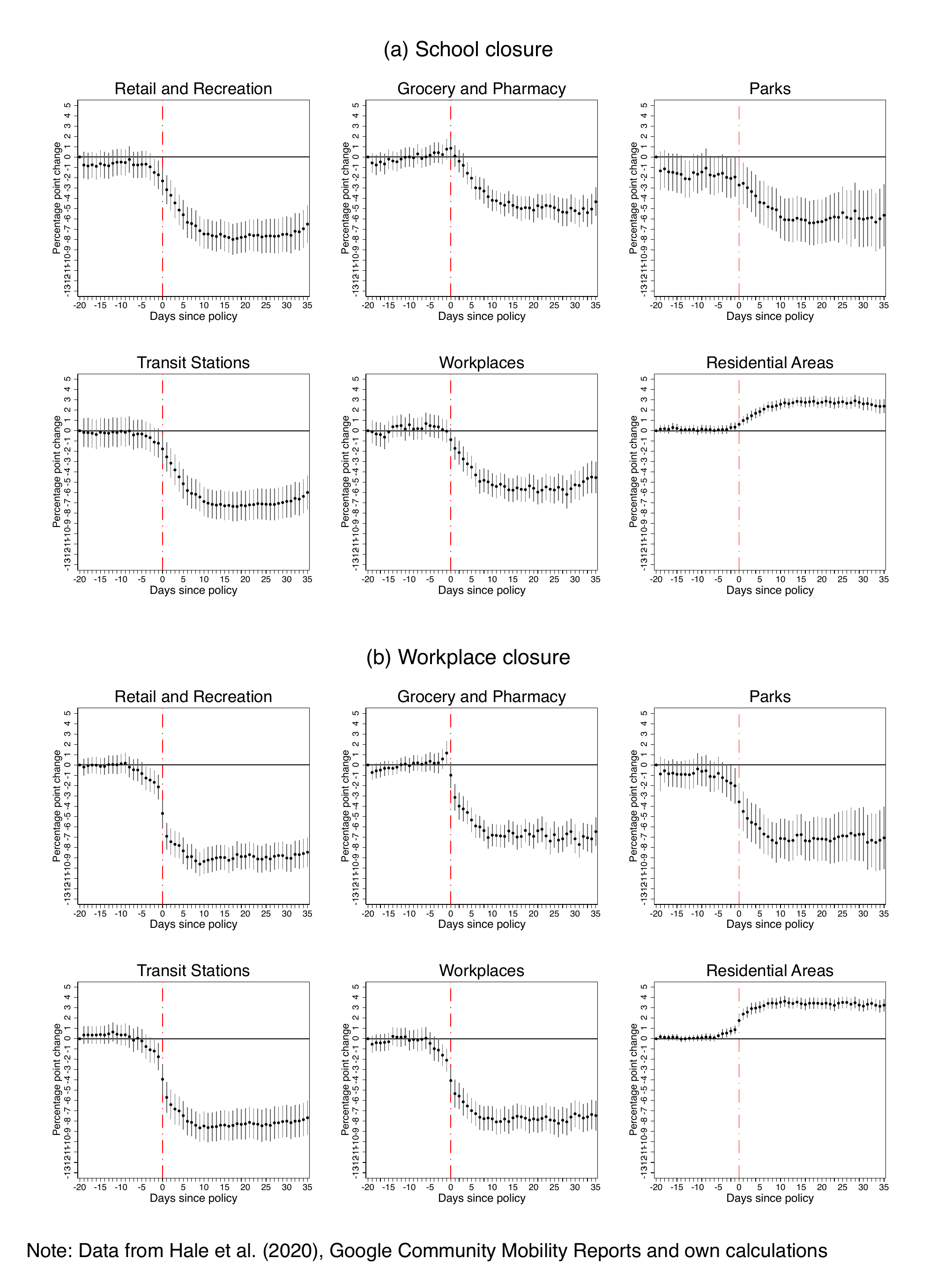}
	\caption{Effects of  {\bf school}  (panel a) and {\bf workplace} (panel b) closures on Google mobility patterns without concurrent policy controls. }
	\label{fig:school_and_workplace_i_single_ss_1_cc_1}
\end{figure}

\newpage


\begin{figure}[!htb]
	\centering
 \includegraphics[width=.85\textwidth]{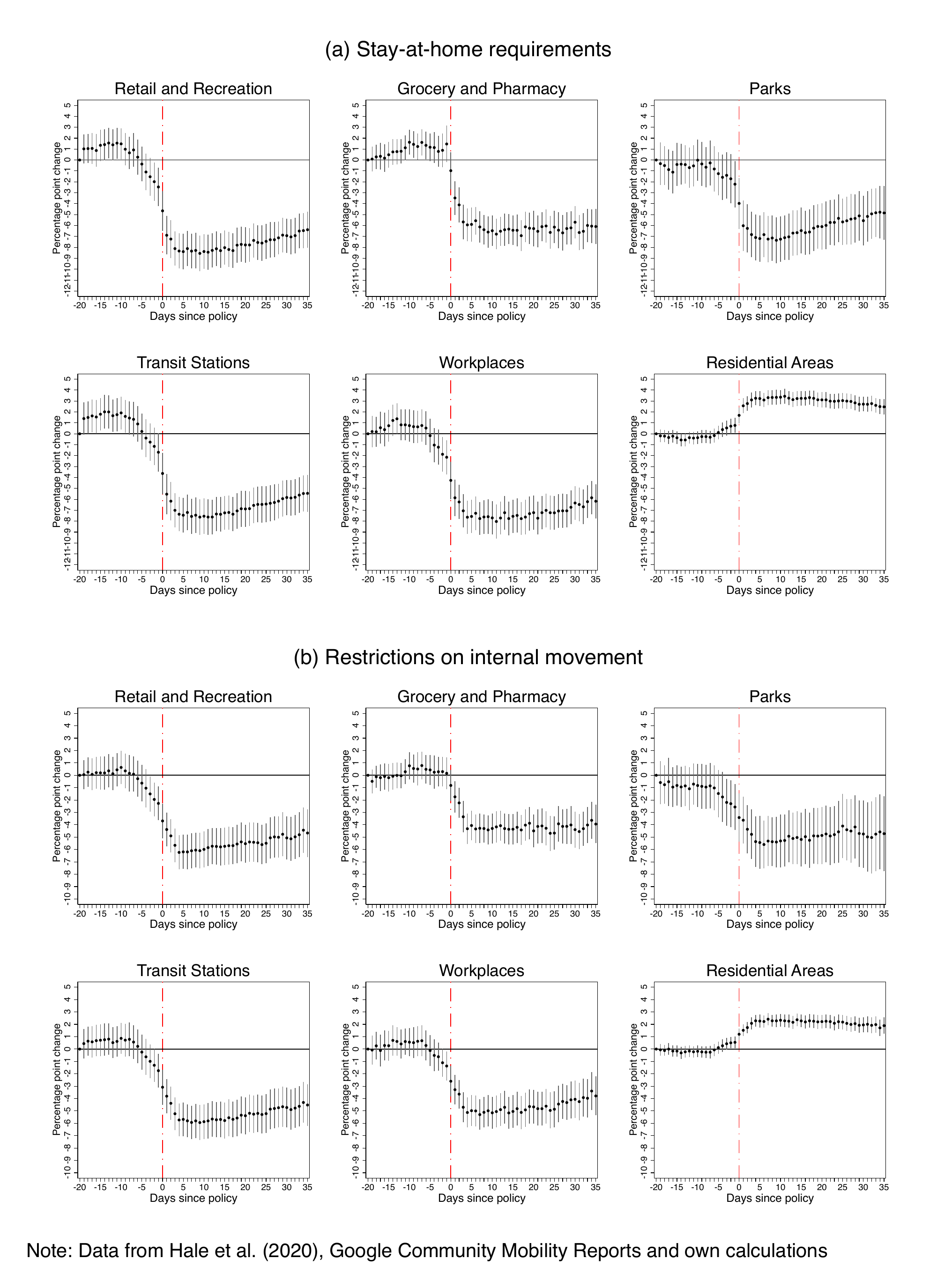}
	\caption{Effects of {\bf stay-at-home requirements} (panel a) and {\bf restrictions on internal mobility} (panel b) on Google mobility patterns without concurrent policy controls. }
	\label{fig:home_and_mobility_i_single_ss_1_cc_1}
\end{figure}

\newpage

\section{Appendix - Sample of countries} \label{appendix:sample}
\begin{description} \footnotesize
\item[Estimations for COVID-19 cases are based on a sample of 135 countries]
\end{description}

\begin{itemize}

\item \noindent \footnotesize
Afghanistan, Angola, Argentina, Australia, Austria, Bahrain, Bangladesh, Barbados, Belgium, Belize, Bolivia, Bosnia and Herzegovina, Botswana, Brazil, Bulgaria, Burkina Faso, Cameroon, Canada, Cape Verde, Chile, Colombia, Costa Rica, Croatia, Czech Republic, Denmark, Dominican Republic, Ecuador, Egypt, El Salvador, Estonia, Finland, France, Gabon, Germany, Ghana, Greece, Guatemala, Honduras, Hong Kong, Hungary, India, Indonesia, Iraq, Ireland, Israel, Italy, Jamaica, Japan, Jordan, Kazakhstan, Kenya, Kuwait, Kyrgyz Republic, Laos, Lebanon, Libya, Luxembourg, Malaysia, Mali, Mauritius, Mexico, Moldova, Mongolia, Mozambique, Myanmar, Namibia, Netherlands, New Zealand, Nicaragua, Niger, Nigeria, Norway, Oman, Pakistan, Panama, Papua New Guinea, Paraguay, Peru, Philippines, Poland, Portugal, Puerto Rico, Qatar, Romania, Rwanda, Saudi Arabia, Serbia, Singapore, Slovak Republic, Slovenia, South Africa, South Korea, Spain, Sri Lanka, Sweden, Switzerland, Tanzania, Thailand, Trinidad and Tobago, Turkey, Uganda, United Arab Emirates, United Kingdom, United States, Uruguay, Vietnam, Zambia, Zimbabwe.
\item \noindent No school intervention: Burundi, Nicaragua.
\item \noindent No workplace intervention: Brunei, Burundi, Eswatini, Mozambique, Nicaragua, Niger, Tanzania.
\item \noindent No events intervention: Burundi, Nicaragua, Sweden.
\item \noindent No transport intervention: Australia, Brunei, Bulgaria, Burundi, Canada, Chile, Czech Republic, Dominica, Estonia, Germany, Hong Kong, Iceland, Japan, Malawi, Malaysia, Mali, Mauritania, Mozambique, Namibia, Nicaragua, Niger, Panama, South Korea, Sweden, Switzerland, Tanzania, Zambia.
\item \noindent No mobility intervention: Burundi, Hong Kong, Iceland, Malawi, Mozambique, Nicaragua, Tanzania.
\item \noindent No travel intervention: Luxembourg, United Kingdom.
\item \noindent No home intervention: Brunei, Burundi, Cameroon, Iceland, Nicaragua, Norway, Sweden, Tanzania.
\end{itemize}

\begin{description} \footnotesize
\item[Estimations for Google Mobility are based on a sample without these 27 countries]
\end{description}

\begin{itemize}

\item \noindent \footnotesize
Albania, Algeria, Azerbaijan, Brunei, Burundi, Chad, China, Cyprus, Democratic Republic of Congo, Dominica, Eswatini, Ethiopia, Gambia, Guyana, Iceland, Lesotho, Madagascar, Malawi, Mauritania, Morocco, Palestine, Russia, Seychelles, Sierra Leone, Tunisia, Ukraine, Uzbekistan.
\end{itemize}

\end{document}